\documentclass[intlimits,twoside,a4paper]{article}

\usepackage[cp1251]{inputenc}
\usepackage[eqsecnum]{cmpj3}

\usepackage{bm}
\usepackage{multicol}

\newcommand{\be}{\begin{equation}}
\newcommand{\ee}{\end{equation}}
\newcommand{\bea}{\begin{eqnarray}}
\newcommand{\eea}{\end{eqnarray}}

\issue{2020}{23}{3}{33702}
\doinumber{10.5488/CMP.23.33702}

\title[Effect of hydrostatic pressure and longitudinal electric field]%
{Effect of hydrostatic pressure and longitudinal electric field on phase transitions and thermodynamic characteristics of quasione-dimensional  CsH$_2$PO$_4$ ferroelectric}

\author[A.S. Vdovych, I.R. Zachek, R.R. Levitskii]{A.S. Vdovych\refaddr{label1}, I.R. Zachek\refaddr{label2}, R.R. Levitskii\refaddr{label1}}
\addresses{
\addr{label1} Institute for Condensed Matter Physics of the National
Academy of Sciences of Ukraine,\\ 1 Svientsitskii St., 79011 Lviv,
Ukraine
\addr{label2} Lviv Polytechnic National University, 1 S. Bandera St., 79013, Lviv, Ukraine}
\date{Received December  12, 2019, in final form March 9, 2020}
\begin{document}

\maketitle

\begin{abstract}
We propose a two-sublattice proton ordering model for the  quasione-dimensional CsH$_2$PO$_4$ ferroelectric with hydrogen bonds, which takes into account
linear on lattice strains  $u_1$, $u_2$, $u_3$ and $u_5$ contribution to the energy of proton subsystem. The model also takes into account the dependence of effective dipole moments of pseudospins on the order parameters, which enables one to agree the  effective dipole moments in paraelectric and ferroelectric phases. Within this model in two-particle cluster approximation on short-range interactions and in the mean field approximation on long-range interactions, there is investigated the behaviour of spontaneous polarization, longitudinal dielectric permittivity and molar heat capacity under the action of hydrostatic pressure and longitudinal electric field. The phase transition into antiferroelectric phase under high pressures is explained. The character of smearing of the paraelectric-ferroelectric phase transition as well as suppression of the antiferroelectric phase at the presence of electric field is studied. 
\end{abstract}

\vspace{-0.7cm}
\section{Introduction}

Investigations of the pressure and field effects on physical properties of ferroelectrics are urgent indeed because they make it possible to deeper understand the mechanisms of phase transitions in these materials as well as to carry out a search for the new physical effects, which are not observed under zero pressure and zero external field.

A ferroelectric with hydrogen bonds CsH$_2$PO$_4$ (CDP) is an example of the crystal where the pressure and field effects are essential. In this crystal there are two structurally inequivalent types of hydrogen bonds of different length (figure~\ref{CDP_ferro_ab} a). Longer bonds have one equilibrium position for protons, whereas shorter bonds have two equilibrium positions. They link PO$_4$ groups into chains along $b$-axis (figure~\ref{CDP_ferro_ab} b); therefore, the crystal is quasione-dimentional. At room temperature and under zero pressure, the crystal is in the paraelectric phase and has a monoclinic symmetry  (space group P2$_{1}$/m) \cite{Matsunaga2011,Itoh2626}. In this case, protons on the shorter bonds occupy two equilibrium positions with equal probability. Below  $T_{c}=153$~K, the crystal passes into the ferroelectric phase (space group P2$_{1}$) \cite{Iwata304,Iwata4044}  with spontaneous polarization along crystallographic  $b$-axis, and the protons occupy mainly the upper equilibrium position (figure~\ref{CDP_ferro_ab}, b).

Based on dielectric investigations \cite{Yasuda1311,Yasuda2755}, it was determined that under pressure $p_{c} =0.33$~GPa and $T_{c}^{\text{cr}}=124.6$~K,  there appear double hysteresis loops, which means that the crystal passes to the antiferroelectric phase. Using investigations on neutron scattering \cite{Schuele935}, it was established that in the antiferroelectric phase a primitive cell of CDP crystal redoubles along  $a$-axis, because there appear two sublattices in the form of (\textit{b,c})-planes, which are polarized antiparallelly along  $b$-axis and alternate along  \textit{a}-axis. Symmetry of the crystal remains monoclinic  (space group P2$_{1}$), and parameters of the redoubled lattice are as follows: $a = 15.625$~\AA, $b = 6.254$~\AA, $c = 4.886$~\AA, $\beta = 108.08^{\,\circ}$. Thus, there takes place a quite large shift of  Cs atoms and  PO$_4$ groups in $(a,c)$-plane and the rotation of the  PO$_4$ groups by 36.8$^{\,\circ}$ around  $b$-axis, which passes through \textit{P} atom. The protons on the hydrogen bonds of the neighbour sublattices are ordered antiparallelly. Under very high pressures, there appears a new antiferroelectric phase  (AF2), where two sublattices have the form of chains along  \textit{b}-axis, which are polarized antiparallelly along $b$-axis and have a checkerboard arrangement. The phase AF2 was predicted in \cite{Schuele2549} based on the NMR investigations  and was confirmed in \cite{Kobayashi83} on the basis of dielectric constant measurements at low temperature and X-ray diffraction measurements.

Results of  dielectric constant measurements under hydrostatic pressure, presented in  \cite{Yasuda2755, Brandt,Kobayashi83,Magome2010}, do not agree with each other. In particular, there are different rates of change of the phase transition temperature with pressure, as well as different maximum values of dielectric permittivity. This is indicative of  high sensitivity of the dielectric properties to the quality of the grown samples of CDP.

An attempt to theoretically describe the paraelectric-ferroelectric and paraelectric-antiferro\-electric phase transition in CsH$_2$PO$_4$ and CsD$_2$PO$_4$ as well as experimental data for the dielectric permittivity was made in  \cite{Blinc6031}, where the crystal is described as pseudospin Ising chains.  The interactions between the pseudospins within a chain are taken into account strictly, whereas the dipole-dipole interactions between the pseudospins of different chains are determined in the mean field approximation. Therein, expressions for spontaneous polarization and dielectric permittivity and equations for the phase transition temperatures were obtained. It was considered that the interactions  linearly decrease with pressure, while the interchain interactions change their sign under the pressure higher than the critical value.
However, there was not examined the issue about a description of the experimental data for dielectric constant  by the proposed theory.

Later on, in \cite{914R} using the pseudospin model \cite{Blinc6031} and the two-particle cluster approximation for short-range configuration interactions,  there were calculated thermodynamic and dynamic characteristics of CDP type ferroelectrics at different values of hydrostatic pressure.  A good agreement of the theory with the experimental data for the dielectric constant and for pressure dependence of the para-ferroelectric and para-antiferroelectric phase transition temperatures was obtained. However, in the model \cite{914R,Blinc6031}, one cannot calculate piezoelectric and elastic characteristics of the crystal, and the critical pressure does not depend on temperature.

In \cite{Deguchi3074}, temperature dependences of the lattice strains $u_1$, $u_2$, $u_3$, $u_5$ were measured. There was also proposed a quasione-dimensional Ising model for  CDP crystal, where the parameters of interaction are linear functions of these strains. Based on this model, the temperature dependences $u_j(T)$ were explained. However, this model does not consider the crystal as two sublattices and does not enable one to describe the ferroelectric-antiferroelectric phase transition under high pressures.

In \cite{FXTT40}, there  was proposed a two-sublattice model of compressible CDP crystal, where the interactions between the neighbouring pseudospins within a chain are taken into account in the two-particle cluster approximation, whereas long-range  (including interchain) interactions --- in the mean field approximation. Here, interaction parameters are linear functions of strains  $u_j$. As a result, there were calculated the temperature dependences  of spontaneous polarization, dielectric permittivity, piezoelectric coefficients and elastic constants.

In \cite{Levitskii4702}, using the proposed in  \cite{FXTT40} model of deformed  CDP crystal, there was investigated the effect of hydrostatic pressure on the phase transition temperature, longitudinal static dielectric characteristics of  Cs(H$_{1-x}$D$_x)_2$PO$_4$ crystals.

As it is well known, at the presence of the longitudinal field $E_y$, a second order phase transition smears, and the temperature dependence of the longitudinal permittivity  $\varepsilon_{yy}(T)$ shows a rounded maximum. At the same time, in \cite{FXTT40,Levitskii4702}, the effective dipole moments, which have different values in the paraelectric and ferroelectric phases, were used to calculate  the longitudinal dielectric permittivity $\varepsilon_{yy}$. This leads to the appearance of a break on the curve  $\varepsilon_{yy}(T)$ instead of the rounded maximum at the presence of the external field  $E_y$. Therefore, in order to describe the smearing of the phase transition, in the present paper we  modified the model \cite{FXTT40} assuming that the effective dipole moment on a hydrogen bond depends on the order parameter on this bond because the order parameter depends on temperature continuously near the phase transition point.

\section{The model of CDP crystal}

\begin{figure}[!t]
	\begin{center}
\includegraphics[scale=0.53]{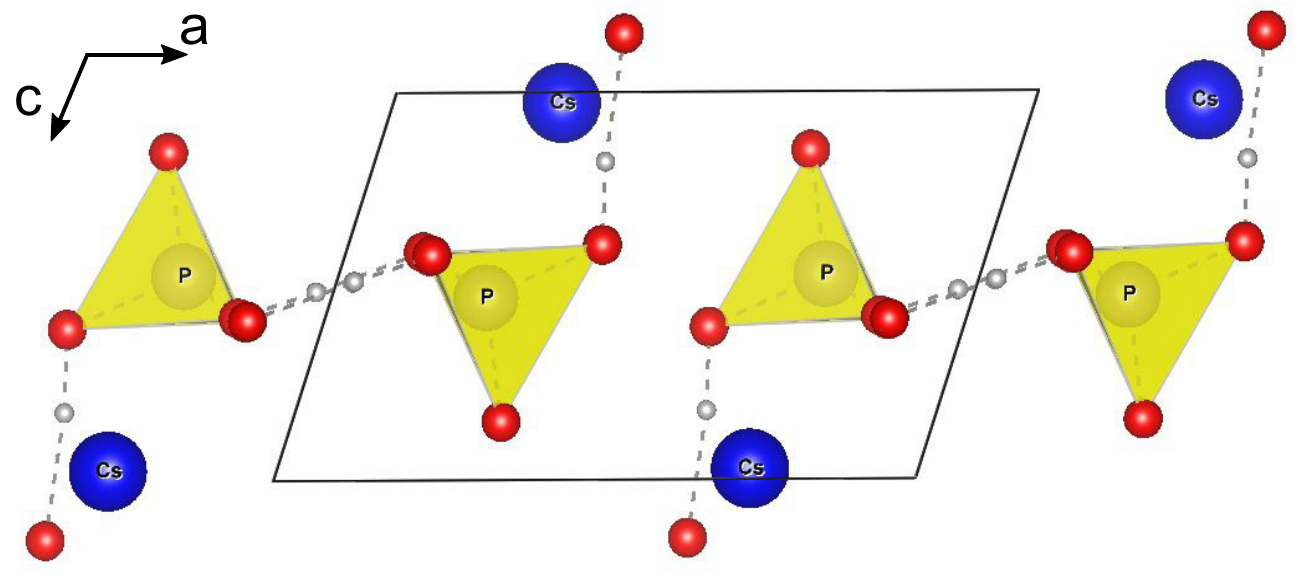} 
\includegraphics[scale=0.53]{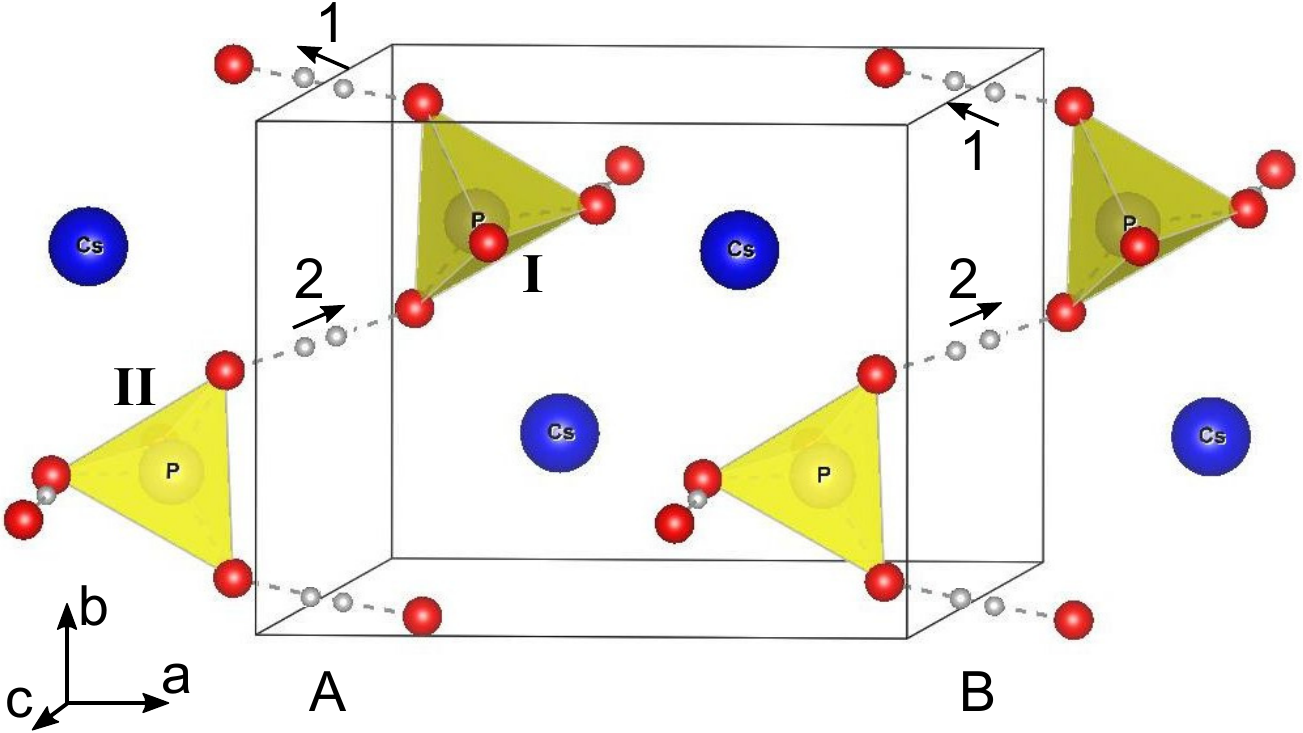}\\
		a	~~~~~~~~~~~~~~~~~~~~~~~~~~~~~~~~~~~~~~~~~~~~~~~~~~~~~~~~~~~~~~~~~	b
	\end{center}
	\caption[]{(Colour online) Primitive cell of CsH$_2$PO$_4$ crystal in the ferroelectric phase.} \label{CDP_ferro_ab}
\end{figure}

We consider the system of protons in CDP, localized on short O-H...O bonds between the groups PO$_{4}$, which form zigzag chains along the crystallographic $b$-axis of the  crystal (see figure~\ref{CDP_ferro_ab}).
The primitive cell includes one chain, marked as ``A'' in figure~\ref{CDP_ferro_ab}. Further, we consider the restricted primitive cell, which includes two chains (``A'' and ``B'') in order to describe the transition to the antiferroelectric phase under high pressure. All the chains ``A'' form a sublattice  ``A'', whereas all the chains  ``B'' form a sublattice ``B''.  Every chain in the primitive cell includes two neighbouring tetrahedra PO$_4$ (of type ``I'' and ``II'') together with two short hydrogen bonds  (respectively, ``1'' and ``2'').

Dipole moments ${\vec{d}}_{q1}^A$, ${\vec{d}}_{q2}^A$, ${\vec{d}}_{q1}^B$, ${\vec{d}}_{q2}^B$, are ascribed to the protons on the bonds.  
Pseudospin variables
$\frac{\sigma_{q1}^A}{2}$, $\frac{\sigma_{q2}^A}{2}$, $\frac{\sigma_{q1}^B}{2}$, $\frac{\sigma_{q2}^B}{2}$  describe  reorientation of the respective dipole moments of the base units: ${\vec{d}}_{q1,2}^{A,B} = \vec\mu_{q1,2}^{A,B} \frac{\sigma_{q1,2}^{A,B}}{2}$.
Mean values  $\langle \frac{\sigma}{2}\rangle = \frac12
(n_a-n_b)$ are connected with differences in occupancy of the two possible molecular positions, $n_a$ and $n_b$.

Hamiltonian of proton subsystem of CDP takes into account short-range and long-range interactions. Under the stresses maintaining the symmetry of crystal  $\sigma_1 = \sigma_{xx}$, $\sigma_2 = \sigma_{yy}$, $\sigma_3 = \sigma_{zz}$,  $\sigma_5 = \sigma_{xz}$ (X $\perp$ (\textit{b},\textit{c}), Y $\parallel$ \textit{b}, Z $\parallel$ \textit{c}), and in the presence of electric field   $E_2=E_y$, it can by written in such a way: 
\setcounter{equation}{0}
\renewcommand{\theequation}{2.\arabic{equation}}
\bea
&& \hat H= NU_{ \text{seed}} + \hat H_{\text{ short}} + \hat H_{\text{long}} + \hat H_{E} + \hat H'_{E}, \label{H}
\eea
where $N$  is the total number of restricted primitive cells.

The first term in (\ref{H})  is ``seed'' energy, which relates  to the heavy ion sublattice and does not explicitly depend on the configuration of the proton subsystem. 
It includes elastic, piezolectric and dielectric parts, expressed in terms of the electric field
$E_2$  and strains maintaining the symmetry of crystal $u_1=u_{xx}$, $u_2=u_{yy}$, $u_3=u_{zz}$, $u_5=2u_{xz}$:
\bea
&& U_{\text{seed}} = v\left( \frac12 \sum\limits_{j,j'} c_{jj'}^{E0} u_ju_j'   - \sum\limits_{j} e_{2j}^0E_2u_j   - \frac12 \chi_{22}^{u 0}E_2^2 \right)  , ~~~~ j, j'=1,2,3,5.  \label{Useed}
\eea
Parameters  $c_{jj'}^{E0}$, $e_{2j}^0$, $\chi_{22}^{u 0}$ are the so-called ``seed'' elastic constants,  ``seed'' coefficients of
piezoelectric stresses and  ``seed'' dielectric susceptibility, respectively; $v$ is the volume of a restricted primitive cell. 
In the paraelectric phase all coefficients $e_{ij}^0 \equiv 0$.

Other terms in  (\ref{H}) describe the pseudospin part of Hamiltonian.
In particular, the second term in  (\ref{H}) is Hamiltonian of short-range interactions:
\be
\hat H_{\text{short}} = - 2w \sum\limits_{qq'} \left ( \frac{\sigma_{q1}^{A}}{2} \frac{\sigma_{q'2}^{A}}{2} + \frac{\sigma_{q1}^{B}}{2}\frac{\sigma_{q'2}^{B}}{2} \right) \left(\delta_{{\bf R}_q{\bf R}_{q'}} + \delta_{{\bf R}_q + {\bf R}_b,{\bf R}_{q'}} \right) . \label{Hshort}
\ee
In (\ref{Hshort}),  $\sigma_{q1,2}^{\text A,B}$ are  $z$-components of pseudospin operator that describe the state of the bond ``1''  or ``2'' of the chain ``A'' or ``B'', in the  $q$-th cell, ${\bf R}_b$ is the lattice vector along OY-axis.
The first Kronecker delta corresponds to the interaction between protons in the chains near the tetrahedra PO$_{4}$ of type ``I'', where the second  Kronecker delta corresponds to the interaction near the tetrahedra PO$_{4}$ of type ``II''. 
Contributions into the energy of interactions between  pseudospins near tetrahedra of different type are identical.

Parameter $w$,  which describes the short-range interactions within the chains, is expanded
linearly into a series with respect to strains $u_j$:
\be
w = w^0 + \sum\limits_{j} \delta_{2j}u_j,\quad(j=1,2,3,5). \label{w}
\ee

The  term $\hat H_{\text{long}}$ in  (\ref{H}) describes  long-range dipole-dipole interactions and indirect  (i.e., through the lattice vibrations)  interactions between protons which are taken into account in the mean field approximation:
\bea
 \hat H_{\text{long}} &=& \frac12 \sum\limits_{{qq'}\atop{ff'}}  J_{ff'}(qq') \frac{\langle \sigma_{qf}^{\text{A}}\rangle}{2}\frac{\langle \sigma_{q'f'}^{\text{A}}\rangle}{2}
- \sum\limits_{{qq'}\atop{ff'}}  J_{ff'}(qq') \frac{\langle \sigma_{q'f'}^{\text{A}}\rangle}{2}\frac{\sigma_{qf}^{\text{A}}}{2} \nonumber \\
& +& \frac12 \sum\limits_{{qq'}\atop{ff'}}  J_{ff'}(qq') \frac{\langle \sigma_{qf}^{\text{B}}\rangle}{2}\frac{\langle \sigma_{q'f'}^{\text{B}}
\rangle}{2}
- \sum\limits_{{qq'}\atop{ff'}}  J_{ff'}(qq') \frac{\langle \sigma_{q'f'}^{\text{B}}\rangle}{2}\frac{\sigma_{qf}^{\text{B}}}{2} \nonumber \\
& +& \frac12 \sum\limits_{{qq'}\atop{ff'}}  K_{ff'}(qq') \frac{\langle \sigma_{qf}^{\text{A}}\rangle}{2}\frac{\langle \sigma_{q'f'}^{\text{B}}\rangle}{2} -
\sum\limits_{{qq'}\atop{ff'}}  K_{ff'}(qq') \frac{\langle \sigma_{q'f'}^{\text{B}}\rangle}{2}\frac{\sigma_{qf}^{\text{A}}}{2} \nonumber \\
& +& \frac12 \sum\limits_{{qq'}\atop{ff'}}  K_{ff'}(qq') \frac{\langle \sigma_{qf}^{\text{B}}\rangle}{2}\frac{\langle \sigma_{q'f'}^{\text{A}}\rangle}{2} -
\sum\limits_{{qq'}\atop{ff'}}  K_{ff'}(qq') \frac{\langle \sigma_{q'f'}^{\text{A}}\rangle}{2}\frac{\sigma_{qf}^{\text{B}}}{2}, \label{Hlong} 
\eea
where the first two terms describe the effective long-range interaction between pseudospins within the same sublattice ``{A}'' or ``{B}'', whereas two other terms  --- between pseudospins of different sublattices ``{A}'' and ``{B}''. Taking into account that  $\langle \sigma_{qf}^{\text{A,B}}\rangle$ does not depend on the number of primitive cells  \textit{q}, we write (\ref{Hlong}) in such a way:
\begin{equation}
\hat H_{\text{long}} = N H^0 + \hat H_2, \label{Hlongg}
\end{equation}
where
\bea
 H^0 &=& \frac18  \left[  J_{11} \left(  \langle \sigma_{1}^{\text{A}} \rangle^2 + \langle \sigma_{1}^{\text{B}} \rangle^2\right)  +J_{22} \left(  \langle \sigma_{2}^{\text{A}} \rangle^2 + \langle \sigma_{2}^{\text{B}} \rangle^2\right)  +  2J_{12} \left(  \langle \sigma_{1}^{\text{A}}\rangle\langle \sigma_{2}^{\text{A}}\rangle  + \langle \sigma_{1}^{\text{B}}\rangle\langle \sigma_{2}^{\text{B}}\rangle\right) \right] \nonumber\\
& +& \frac{1}{4}\left[  K_{11}\langle \sigma_{1}^{\text{A}} \rangle \langle \sigma_{1}^{\text{B}} \rangle + K_{22}\langle \sigma_{2}^{\text{A}} \rangle \langle \sigma_{2}^{\text{B}} \rangle+ K_{12}\left( \langle \sigma_{1}^{\text{A}} \rangle \langle \sigma_{2}^{\text{B}} \rangle +\langle \sigma_{1}^{\text{B}} \rangle \langle \sigma_{2}^{\text{A}} \rangle\right)  \right]  ,\label{H0}  
\eea
\bea
 \hat H_2 =&-&\sum\limits_{q}\left\lbrace  \left( J_{11}\langle \sigma_{1}^{\text{A}}\rangle+J_{12}\langle \sigma_{2}^{\text{A}}\rangle+K_{11}\langle \sigma_{1}^{\text{B}}\rangle + K_{12}\langle \sigma_{2}^{\text{B}}\rangle\right) \frac{\sigma_{q1}^{A}}{4}\right.\nonumber\\ 
 &+&\left( J_{12}\langle \sigma_{1}^{\text{A}}\rangle+J_{22}\langle \sigma_{2}^{\text{A}}\rangle+K_{12}\langle \sigma_{1}^{\text{B}}\rangle + K_{22}\langle \sigma_{2}^{\text{B}}\rangle\right) \frac{\sigma_{q2}^{A}}{4} \nonumber\\
& +&\left( J_{11}\langle \sigma_{1}^{\text{B}}\rangle+J_{12}\langle \sigma_{2}^{\text{B}}\rangle+K_{11}\langle \sigma_{1}^{\text{A}}\rangle + K_{12}\langle \sigma_{2}^{\text{A}}\rangle\right) \frac{\sigma_{q1}^{B}}{4} \nonumber\\
&+&\left.\left( J_{12}\langle \sigma_{1}^{\text{B}}\rangle+J_{22}\langle \sigma_{2}^{\text{B}}\rangle+K_{12}\langle \sigma_{1}^{\text{A}}\rangle + K_{22}\langle \sigma_{2}^{\text{A}}\rangle\right) \frac{\sigma_{q2}^{B}}{4}\right\rbrace . \label{H2}
\eea
Here, parameters $J_{ff'} = \sum\nolimits_{{\bf R}_q-{\bf R}_{q'}}J_{ff'}(qq')$ and $K_{ff'} = \sum\nolimits_{{\bf R}_q-{\bf R}_{q'}}K_{ff'}(qq')$ are Fourier transforms of long-range interaction constants  at ${\bf k}=0$. 
The parameters $J_{ff'}$ and $K_{ff'}$  are expanded
linearly into a series with respect to strains $u_j$:
\bea
&& J_{11}= J_{22} \!= \! J_{1} +\! \sum\limits_{j} \bar\varphi_{1j}u_j,~~~~~~~~~  J_{12}= \!J_{21}=\! J_{2}\! + \!\sum\limits_{j} \bar\varphi_{2j}u_j, \nonumber \\
&& K_{11}= K_{22} = K_1 +  \sum\limits_{j}\varphi_{1j}u_j,~~~~\ K_{12} =K_{21} = K_2 + \sum\limits_{j} \varphi_{2j}u_j.
\eea
Taking into account such symmetry of pseudospins in the chains of  CDP
\be
\langle \sigma_{q1}^{\text{A}} \rangle = \langle \sigma_{q2}^{\text{A}} \rangle = \eta_{1}, \quad
\langle \sigma_{q1}^{\text{B}} \rangle = \langle \sigma_{q2}^{\text{B}} \rangle = \eta_{2}, \label{eta_12}
\ee
we write expressions (\ref{H0}), (\ref{H2}) in such a way:
\bea
&& \hat H^{0} = \nu_{1}( \eta_{1}^2 + \eta_{2}^2) + 2\nu_{2}\eta_{1}\eta_{2}, 
\eea
\bea
&& \hat H_{2} = \sum\limits_{q} \left[  - \left( 2\nu_{1}\eta_{1} +  2 \nu_{2}\eta_{2} \right)   \left( \frac{\sigma_{q1}^{A}}{2} + \frac{\sigma_{q2}^{A}}{2} \right) - (2\nu_{2}\eta_{1} +  2 \nu_{1}\eta_{2} )   \left( \frac{\sigma_{q1}^{B}}{2} + \frac{\sigma_{q2}^{B}}{2} \right)\right]  .
\eea
Here, the following notations are used:
\bea
&& \hspace{-8ex}  \nu_{1} = \frac{1}{8}(J_{11}+J_{22}+2J_{12})= \nu_{1}^0 + \sum\limits_{j}  \psi_{j1}u_j,\quad\,\,\nu_{1}^0=\frac14(J_{1} + J_2),\quad\,\,\,\, \psi_{j1}=\frac14(\bar\varphi_{1j} + \varphi_{1j}), \label{nu1}\\
&& \hspace{-8ex}  \nu_{2} = \frac{1}{8}(K_{11}+K_{22}+2K_{12})= \nu_{2}^0 + \sum\limits_{j}  \psi_{j2}u_j,\quad\nu_{2}^0=\frac14(K_{1} + K_2), \quad \psi_{j2}=\frac14(\bar\varphi_{2j} + \varphi_{2j}). \label{nu2}
\eea

The fourth term in  (\ref{H}) describes the interactions of pseudospins with the external electric field:
\bea
&&\hat H_{E}= - \sum\limits_q \mu_y E_2 \left( \frac{\sigma_{q1}^{\text{A}}}{2} + \frac{\sigma_{q2}^{\text{A}}}{2} + \frac{\sigma_{q1}^{\text{B}}}{2} + \frac{\sigma_{q2}^{\text{B}}}{2} \right),
\eea
where $\mu_y$ is $y$-component of effective dipole moments per one pseudospin.

The term $\hat H'_{E}$ in Hamiltonian (\ref{H}) takes into account the above mentioned  dependence of effective dipole moments on the mean value of pseudospin $s_f$:
\bea
&&\hat H'_{E} = -\sum\limits_{qf} s_f^2 \mu' E_2 \frac{\sigma_{qf}}{2} = - \sum\limits_{qf} \left(\frac{1}{N}\sum\limits_{q'}\sigma_{q'f}\right)^2 \mu' E_2 \frac{\sigma_{qf}}{2}. \label{H_E}
\eea
where $ \sigma_{qf} $ (\textit{f}=1, 2, 3, 4) are brief notations of pseudospins $\sigma_{q1}^{\text{A}}$, $\sigma_{q2}^{\text{A}}$, $\sigma_{q1}^{\text{B}}$, $\sigma_{q2}^{\text{B}}$, respectively. 
Here, we use corrections to dipole moments $s_f^2 \mu'$ instead of $s_f \mu'$ due to the symmetry considerations, the energy should not change, when field and all pseudospins change their sign.

The term $\hat H'_E$, as well as long-range interactions, are taken into account in the mean field approximation: 
\bea
  \hat H'_{E} &=& - \sum\limits_{qf} \left(\frac{1}{N}\sum\limits_{q'}\sigma_{q'f}\right)^2 \mu' E_2 \frac{\sigma_{qf}}{2} =    - \frac{1}{N^2} \sum\limits_{qf} \sum\limits_{q'} \sum\limits_{q''} \sigma_{qf} \sigma_{q'f} \sigma_{q''f} \frac{\mu'E_2}{2} \nonumber \\
& \approx&  - \frac{1}{N^2} \!\sum\limits_{qf} \!\sum\limits_{q'} \!\sum\limits_{q''}\! \left[  \left( \sigma_{qf} + \sigma_{q'f} + \sigma_{q''f}\right)  \eta_{f}^2 \!-\! 2\eta_{f}^3\right]  \frac{\mu'E_2}{2}\nonumber\\
& =&  - 3 \sum\limits_{q}\sum\limits_{f=1}^4 \frac{\sigma_{qf}}{2} \eta_f^2 \mu'E_2
+ N\sum\limits_{f=1}^4\eta_f^3 \mu' E_2.
\label{H'_E} 
\eea
Taking into account (\ref{eta_12}), expression  (\ref{H'_E}) can be written as:
\bea
&& \hspace{-2ex}  \hat H'_{E} =   -3 \sum\limits_{q} \mu'E_2 \left( \frac{\eta_1^2\sigma_{q1}^{\text{A}}}{2} + \frac{\eta_1^2\sigma_{q2}^{\text{A}}}{2} + \frac{\eta_2^2\sigma_{q1}^{\text{B}}}{2} + \frac{\eta_2^2\sigma_{q2}^{\text{B}}}{2} \right)  + 2N(\eta_1^3+\eta_2^3) \mu'E_2.
\eea

The two-particle cluster approximation for short-range interactions is used to calculate the thermodynamic characteristics of CDP. In this approximation, thermodynamic potential is given by:
\bea
&& \hspace{-4ex} G = N U_{ \text{seed}} + NH^0  + 2N(\eta_1^3+\eta_2^3) \mu'E_2 - N v \sum\limits_{j}  \sigma_j u_j - \nonumber\\
&& - k_\text{B} T \sum\limits_q \left\{ 2 \ln {\rm Sp} \text{e}^{-\beta \hat H^{(2)}_{q}}  \left. -  \ln {\rm Sp} \text{e}^{-\beta \hat H^{(1)\text{A}}_{q}} -  \ln {\rm Sp} \text{e}^{-\beta \hat H^{(1)\text{B}}_{q}} \right\} \right., \label{G}
\eea
where $\beta=\frac{1}{k_\text{B}T}$, $k_\text{B}$ is Boltzmann constant, $\hat H^{(2)}_{q}$, $\hat H^{(1)A}_{q}$, $\hat H^{(1)B}_{q}$ are two-particle and one-particle Hamiltonians:
\bea
&& \hat H^{(2)}_{q} = - 2 w\left( \frac{\sigma_{q1}^{\text{A}}}{2} \frac{\sigma_{q2}^{\text{A}}}{2} +
\frac{\sigma_{q1}^{\text{B}}}{2} \frac{\sigma_{q2}^{\text{B}}}{2}\right) -  \frac{y_{1}}{\beta} \left( \frac{\sigma_{q1}^{\text{A}}}{2} + \frac{\sigma_{q2}^{A}}{2} \right) -
\frac{y_{2}}{\beta} \left( \frac{\sigma_{q1}^{\text{B}}}{2} + \frac{\sigma_{q2}^{\text{B}}}{2} \right), \label{H2q}\\
&& \hat H^{(1)\text{A}}_{q} = - \frac{\bar y_{1}}{\beta} \left( \frac{\sigma_{q1}^{\text{A}}}{2} + \frac{\sigma_{q2}^{\text{A}}}{2} \right), ~~
\hat H^{(1)\text{B}}_{q} = - \frac{\bar y_{2}}{\beta} \left( \frac{\sigma_{q1}^{\text{B}}}{2} + \frac{\sigma_{q2}^{\text{B}}}{2} \right), \label{H1q}
\eea
where such notations are used:
\bea
&& y_{1} = \beta \Delta_{1} + 2\beta \nu_{1}\eta_{1}  +   2\beta \nu_{2} \eta_{2}+ \beta(\mu_yE_2 + 3\eta_1^2 \mu'E_2),  \label{y1}\\
&& y_{2} = \beta \Delta_{2} + 2\beta \nu_{2}\eta_{1}  +   2\beta \nu_{1} \eta_{2}+ \beta(\mu_yE_2 + 3\eta_2^2 \mu'E_2) , \label{y2}\\
&& \bar y_{1} =  \beta \Delta_{1} + y_{1}, ~~ \bar y_{2} =  \beta \Delta_{2} + y_{2}. \nonumber
\eea
Symbols $\Delta_l$  are effective field, created by the neighboring bonds
from outside the cluster. In the cluster approximation, these fields can be determined from the condition of minimum of thermodynamic potential $\partial G/\partial \Delta_l =0$, which gives   the self-consistency
condition, which states that the mean values of the pseudospins $\langle \sigma_{qf}^{\text{A,B}} \rangle$ calculated using two-particle
and one-particle Gibbs distribution, respectively, should coincide;
that is,
\bea
&& \eta_{1} = \frac{{\rm Sp}\, \sigma_{qf}^{\text{A}} \text{e}^{-\beta \hat H^{(2)}_{q}}}{{\rm Sp} \, \text{e}^{-\beta \hat H^{(2)}_{q}}} =
\frac{{\rm Sp} \, \sigma_{qf}^{\text{A}} \text{e}^{-\beta \hat H^{(1)\text{A}}_{q}}}{{\rm Sp} \, \text{e}^{-\beta \hat H^{(1)\text{A}}_{q}}},\nonumber \\
&& \eta_{2} = \frac{{\rm Sp}\, \sigma_{qf}^{\text{B}} \text{e}^{-\beta \hat H^{(2)}_{q}}}{{\rm Sp} \, \text{e}^{-\beta \hat H^{(2)}_{q}}} =
\frac{{\rm Sp} \, \sigma_{qf}^{\text{B}} \text{e}^{-\beta \hat H^{(1)\text{B}}_{q}}}{{\rm Sp} \, \text{e}^{-\beta \hat H^{(1)\text{B}}_{q}}}.
\label{Sp12} 
\eea
Hence, based on   (\ref{Sp12}),  taking into account  (\ref{eta_12}), (\ref{H2q}) and (\ref{H1q}), we obtain expressions for the order parameters:
\bea
&& \eta_{1} = \frac{1}{D} \left[ \sinh (y_{1} + y_{2}) +  \sinh (y_{1} - y_{2}) + 2a \sinh y_{1}\right] = \tanh \frac{\bar y_{1}}{2}, \nonumber \\
&& \eta_{2} = \frac{1}{D} \left[ \sinh (y_{1} + y_{2}) -  \sinh (y_{1} - y_{2}) + 2a \sinh y_{2} \right] = \tanh \frac{\bar y_{2}}{2},\label{eta} 
\eea
where such notations are used:
\bea
&& D = \cosh  (y_{1} + y_{2}) +  \cosh  (y_{1} - y_{2}) + 2a \cosh  y_{1}+ 2a \cosh  y_{2}  + 2a^2,  \nonumber\\
&&a = \text{e}^{-\frac{w}{k_\text{B}T}}.\nonumber
\eea
Excluding the cluster fields $\Delta_{l}$  from expression $\eta_{l} = \tanh (\bar y_l/2)$ [see (\ref{eta})], we  write (\ref{y1}), (\ref{y2})  in such a way:
\bea
&& y_{1} = \frac12 \ln \frac{1 +  \eta_{1}}{1 - \eta_{1}}  + \beta \nu_{1}\eta_{1}  +   \beta \nu_{2} \eta_{2} + \frac{1}{2}\beta\left( \mu_yE_2 + 3\eta_1^2 \mu'E_2\right) , \label{y1x}
\eea
\bea
&& y_{2} = \frac12 \ln \frac{1 +  \eta_{2}}{1 - \eta_{2}}  + \beta \nu_{2}\eta_{1}  +   \beta \nu_{1} \eta_{2}+ \frac{1}{2}\beta\left( \mu_yE_2 + 3\eta_2^2 \mu'E_2\right) . \label{y2x}
\eea
\section {Longitudinal dielectric and thermal characteristics of  CDP}

Using (\ref{G}), thermodynamic potential per one restricted primitive cell  can be written in such a way:
\setcounter{equation}{0}
\renewcommand{\theequation}{3.\arabic{equation}}
\bea
 g &=& U_{ \text{seed}} + H^0  + 2\left( \eta_1^3+\eta_2^3\right)  \mu'E_2 + 2k_\text{B}T \ln 2 - 2w - v \sum\limits_{j} \sigma_j u_j \nonumber  \\
& -&k_\text{B}T\ln \left(  1 -  \eta_{1}^2\right)  - k_\text{B}T\ln ( 1 -  \eta_{2}^2 )  - 2k_\text{B}T\ln D.\label{g}
\eea
Using equilibrium condition
\bea
&& \left( \frac{\partial g}{\partial u_j} \right)_{E_2} = 0,  \nonumber
\eea
we obtain equations for strains  $u_j$:
\bea
&& \hspace{-8ex} \sigma_j = c_{j1}^{E0}u_1 + c_{j2}^{E0}u_2 + c_{j3}^{E0}u_3 + c_{j5}^{E0}u_5 - e_{2j}^0E_2 - \frac{2\delta_j}{v} + \frac{4\delta_j}{vD}M  - \frac{1}{v} \psi_{j1} ( \eta_{1}^{2}  +  \eta_{2}^{2}  ) - \frac{2}{v} \psi_{j2} \eta_{1}\eta_{2},  \label{sigma} 
\eea
where 
\[ M=\bigl[   a\cosh  y_{1} + a\cosh  y_{2} + 2a^{2} \bigr]. \]
In the case of applying the hydrostatic pressure $\sigma_1=\sigma_2=\sigma_3=-p$,  $\sigma_4=\sigma_5=\sigma_6=0$.

Based on  thermodynamic potential (\ref{g}), we get  expressions for different thermodynamic characteristics.
In particular, an expression for longitudinal polarization  $P_2$:
\bea
&& P_2 = -\left( \frac{\partial g}{\partial E_2} \right)_{\sigma_j}   =   \sum\limits_{j} e_{2j}^0u_j   + \chi_{22}^{u 0}E_2  + \frac{\mu_y}{v} \bigl(  \eta_{1} + \eta_{2} \bigr)  +  \frac{\mu'}{v} \bigl(  \eta_{1}^3 + \eta_{2}^3 \bigr). \label{P2}
\eea
Isothermic dielectric susceptibility of a mechanically clamped crystal
is given by:
\bea
\chi_{22}^{u} &=& \left( \frac{\partial P_2}{\partial E_2} \right)_{u_j} = \chi_{22}^{u 0} + \frac{\beta\tilde{\mu}_{1y}^2}{2v\Delta}\left[  D( \varkappa_{11}+\varkappa_{12})-  (\tilde{\varphi}_{2}-\beta\nu_{2})(\varkappa_{11}\varkappa_{22}-\varkappa_{12}^{2})\right]  \nonumber \\ 
&+& \frac{\beta\tilde{\mu}_{2y}^2}{2v\Delta}\left[  D( \varkappa_{12}+\varkappa_{22})-  (\tilde{\varphi}_{1}-\beta\nu_{2})(\varkappa_{11}\varkappa_{22}-\varkappa_{12}^{2})\right]  ,
\label{X22} 
\eea
with the following notations:
\bea
\Delta=D^{2}-D\left[ \tilde{\varphi}_{1}\varkappa_{11}+\tilde{\varphi}_{2}\varkappa_{22}+2\beta\nu_{2}\varkappa_{12}\right] 
+\left[ \tilde{\varphi}_{1}\tilde{\varphi}_{2}-(\beta\nu_{2})^{2}\right] \left( \varkappa_{11}\varkappa_{22}-\varkappa_{12}^{2}\right) ,\nonumber
\eea
\[
\tilde{\varphi}_{1}= \varphi_{1}+3\eta_{1}\beta\mu'E_{2},\quad\tilde{\varphi}_{2}= \varphi_{2}+3\eta_{2}\beta\mu'E_{2}, 
\]
\[
 \varphi_{1} = \frac{1}{1 - \eta_{1}^{2}} + \beta  \nu_{1},\quad \varphi_{2}= \frac{1}{1 - \eta_{2}^{2}} + \beta  \nu_{1},  
\]
\[
 \tilde{\mu}_{1y}=\mu_{y}+3\mu'\eta_{1}^{2},\quad\tilde{\mu}_{2y}=\mu_{y}+3\mu'\eta_{2}^{2},
\] 
\[
 \varkappa_{11}= \cosh (y_{1} + y_{2})+\cosh (y_{1} - y_{2})+2a\cosh y_{1}-\eta_{1}^{2}D,\
 \]
 \[
\varkappa_{12}=\cosh (y_{1} + y_{2})-\cosh (y_{1} - y_{2})-\eta_{1}\eta_{2}D,\]
\[
\varkappa_{21}=\cosh (y_{1} + y_{2})-a^{2}\cosh (y_{1} - y_{2})+2a\cosh y_{2}-\eta_{1}\eta_{2}D,\]
\[
 \varkappa_{22}=\cosh (y_{1} + y_{2})+\cosh (y_{1} - y_{2})+2a\cosh y_{2}-\eta_{2}^{2}D.
\]
%
%

%
%

Molar heat capacity of the proton subsystem of CDP at a constant pressure can be found by numerical differentiation of thermodynamic potential:
\bea
&&\Delta C_p = - \frac{N_AT}{4} \left( \frac{\partial^2 g}{\partial T^2} \right)_{\sigma_j}.
\eea
where $N_\text{A}$ is the Avogadro  constant.

\section{Comparison of theoretical results with the experimental data. Discussion.}

The theory parameters are  determined  from the condition of agreement of the calculated characteristics with experimental data for temperature dependences of spontaneous polarization $P_2(T)$ and dielectric permittivity  $\varepsilon_{22}(T)$ at different values of hydrostatic pressure  \cite{Yasuda2755}, spontaneous strains $u_j$ \cite{Deguchi3074}, molar heat capacity \cite{Imai3960} and elastic constants  \cite{Prawer63};  as well as the agreement with  ab-initio calculations of the  lattice contributions to molar heat capacity  \cite{Shchur301} and dielectric permittivity \cite{VanTroeye024112}.

Parameters of short-range interactions $w_0$ and long-range interactions  $\nu_1^{0}$ (``intra-sublattice''), $\nu_2^{0}$ (``inter-sublattice'') mainly fix the phase transition temperature from  paraelectric to ferroelectric phase at the absence of external pressure and field, the order of  phase transition and the shape of curve $P_2(T)$. Their optimal values are: $w_0/k_{\text{B}}$=650~K,  $\nu_1^{0}/k_{\text{B}}$=1.50~K,  $\nu_2^{0}/k_{\text{B}}$=0.23~K.

To determine the deformational potentials $\delta_{j}$ [see(\ref{w})] and $\psi_{j1}$ (\ref{nu1}), $\psi_{j2}$ (\ref{nu1}) it is necessary to use experimental data for shift of the phase transition temperature under hydrostatic and uniaxial pressures as well as the data for temperature dependences of  spontaneous strains $u_j$, piezoelectric coefficients and elastic constants. Unfortunately,  only data for  spontaneous strains and  hydrostatic pressure effect on the dielectric characteristics are available. 
As a result,  the experimental data for strains and dielectric characteristics can be described  using a great number of  combinations of parameters $\psi_{j1}$,  $\psi_{j2}$. 
Therefore, for the sake of simplicity, we chose $\psi_{j2}$ to be proportional to $\psi_{j1}$.  
Optimal values of deformational potentials are: 
$\delta_{1}/k_{\text{B}}=1214$~K, $\delta_{2}/k_{\text{B}}=454$~K, $\delta_{3}/k_{\text{B}}=1728$~K, $\delta_{5}/k_{\text{B}}=1214$~K,  $\delta_{5}/k_{\text{B}}=-13$~K;    
$\psi_{11}/k_{\text{B}} = 92.2$~K,  $\psi_{21}/k_{\text{B}} = 23.2$~K,  $\psi_{31}/k_{\text{B}} = 139.7$~K,  $\psi_{51}/k_{\text{B}} = 5.5$~K; 
$\psi_{j2} = \frac{1}{3}\psi_{j1}$.

The effective dipole  moment in the paraelectric phase is found  from the condition of agreement of calculated curve $\varepsilon_{22}(T)$  with experimental data. We consider it to be dependent on the value of hydrostatic pressure
\textit{p}, that is $\mu_{y}=\mu_{y}^0(1-k_pp)$, where $\mu_{y}^0=2.63\cdot 10^{-18}$~esu$\cdot$cm, $k_p=0.4\cdot 10^{-10}$~cm$^2$/dyn. 
The correction to the effective dipole moment $\mu'=-0.43\cdot 10^{-18}$~esu$\cdot$cm is found  from the condition of agreement of the calculated saturation polarization with experimental data.

The ``seed'' dielectric susceptibility  $\chi_{22}^{u 0}$, coefficients of piezoelectric stress $e_{2j}^0$ and elastic constants $c_{ij}^{E0}$  are found  from the condition of agreement of theory  with experimental data in the temperature regions far from the phase transition temperature  $T_c$. Their values are obtained as follows:
$\chi_{22}^{u 0}=0.443$~\cite{VanTroeye024112};
$e_{2j}^0=0\,\,{\text{esu}}/{\rm cm^2}$;
$c_{11}^{0E} = 28.83 \cdot 10^{10}$ ${\text{dyn}}/{\rm cm^2}$, $c_{12}^{E0} = 11.4 \cdot 10^{10}$ ${\text{dyn}}/{\rm cm^2}$, $c_{13}^{E0} = 42.87 \cdot 10^{10}$~${\text{dyn}}/{\rm cm^2}$,
$c_{22}^{E0} = 26.67 \cdot 10^{10}$ ${\text{dyn}}/{\rm cm^2}$,
$c_{23}^{E0} = 14.5 \cdot 10^{10}$ ${\text{dyn}}/{\rm cm^2}$, $c_{33}^{E0} = 65.45 \cdot 10^{10}$ ${\text{dyn}}/{\rm cm^2}$, $c_{15}^{E0} = 5.13 \cdot 10^{10}$~${\text{dyn}}/{\rm cm^2}$,  $c_{25}^{E0} = 8.4 \cdot 10^{10}$ ${\text{dyn}}/{\rm cm^2}$, $c_{35}^{E0} = 7.50 \cdot 10^{10}$ ${\text{dyn}}/{\rm cm^2}$,
$c_{55}^{E0} = 5.20 \cdot 10^{10}$~${\text{dyn}}/{\rm cm^2}$.

The volume of a restricted primitive cell is  $\upsilon = 0.467\cdot 10^{-21}$ cm$^3$ \cite{Schuele935}.

\begin{figure}[!t]
	\begin{center}
\includegraphics[scale=0.8]{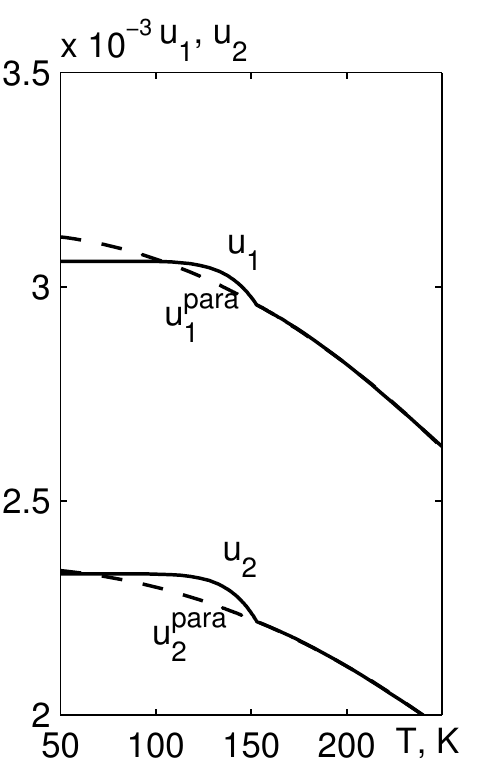} 
\includegraphics[scale=0.8]{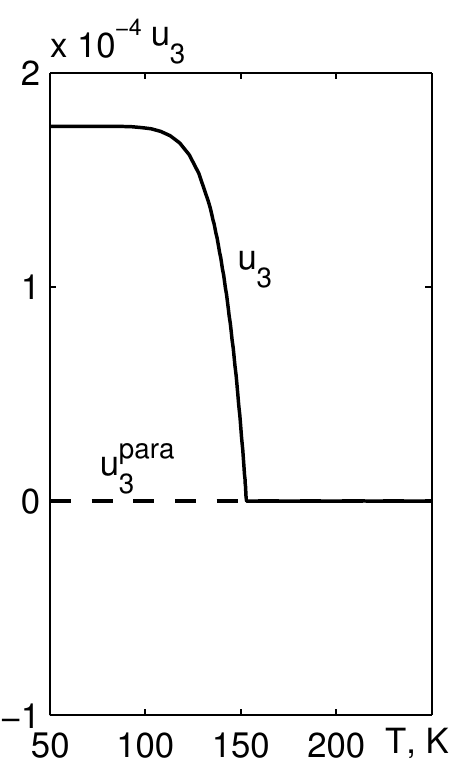}
 \includegraphics[scale=0.8]{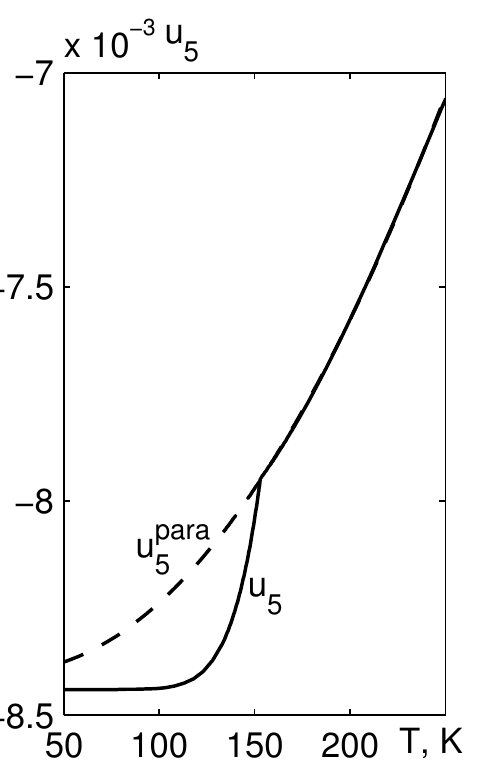}\\
	\end{center}
	\caption[]{Temperature dependence of lattice  strains  $u_j$ under zero pressure.} \label{u_j}
\end{figure}

\begin{figure}[!t]
	\begin{multicols}{2}
	\begin{center}
\includegraphics[scale=0.8]{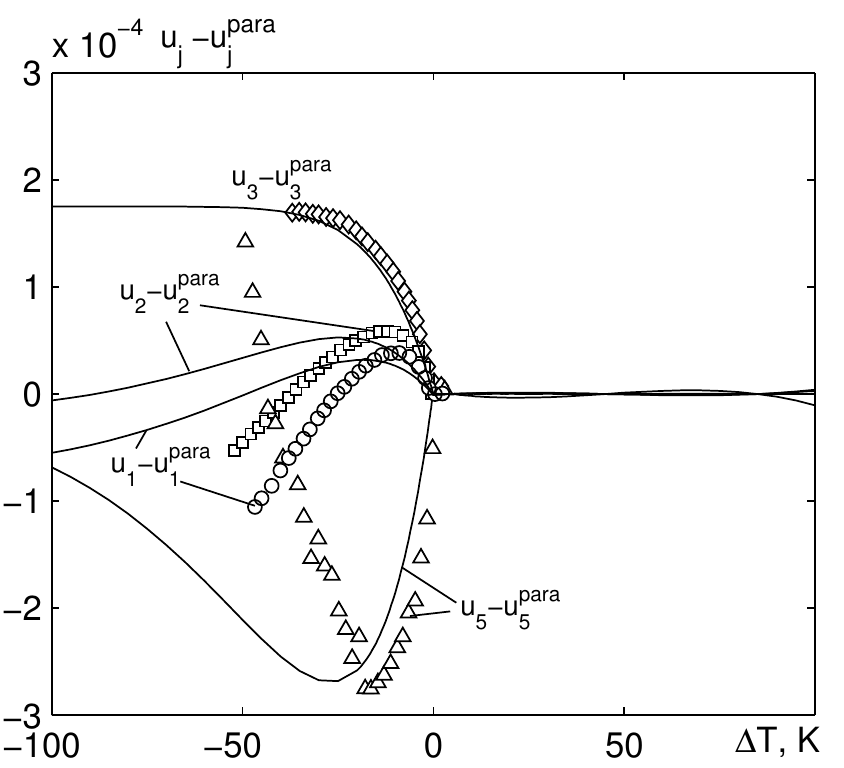} \\
\end{center}
\caption[]{Temperature dependence of differences $u_j-u_j^{\text{para}}$ under zero pressure. Symbols are experimental data  \cite{Deguchi3074}.} \label{Delta_u}
	\begin{center}
\includegraphics[scale=0.8]{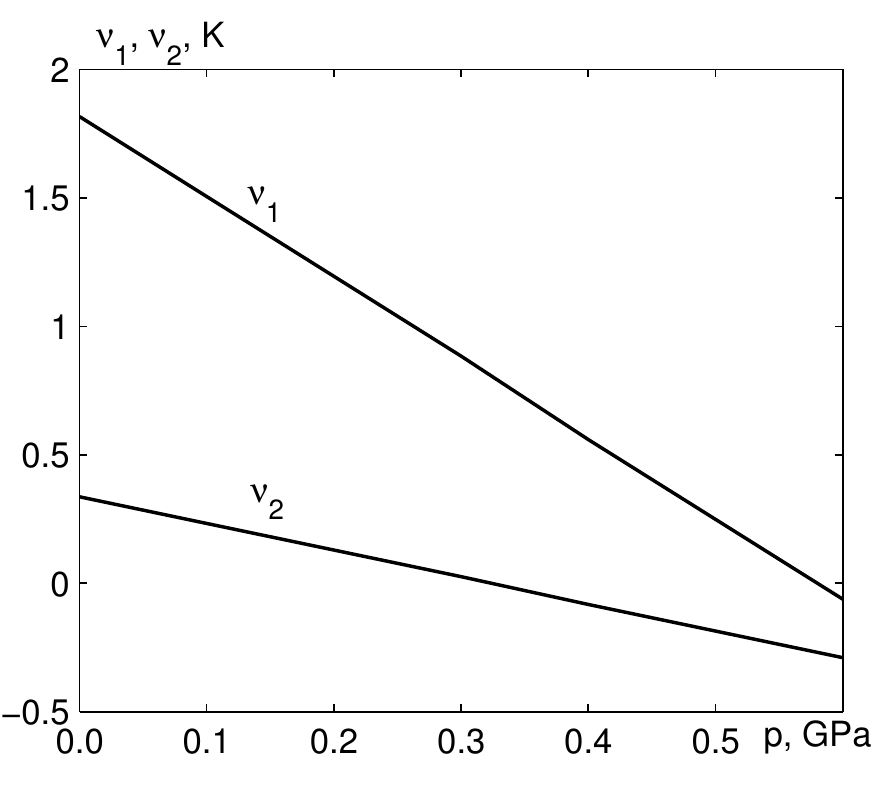}
\end{center}
\caption[]{ Pressure dependence of the parameters of long-range interactions  $\nu_1$, $\nu_2$ at the temperature 80~K.} \label{nu_p}
	\end{multicols}
\end{figure}

Now, let us dwell on the obtained results.
The effect of hydrostatic pressure depend mainly on the behaviour of lattice strains $u_j$ under pressure. 
Temperature dependence of these strains is presented in figure~\ref{u_j}  by solid lines. 
In the paraelectric phase, they depend on temperature almost quadratically  (dashed lines $u_j^{\text{para}}$ in figure~\ref{u_j}), in the temperature range $0<T-T_c<100$~K. Nonzero strains in the paraelectric phase appear as a result of competition of energy of short-range interactions between pseudospins  (\ref{Hshort}) [which depends on strains according to (\ref{w})] and elastic energy $Nv \frac12 \sum\nolimits_{j,j'} c_{jj'}^{E0} u_ju_j'$ [see (\ref{Useed})]. They are an additional contribution to the thermal strains connected with anharmonicity of interatomic interactions, that was also noted  earlier in  \cite{Deguchi3074}. In the ferroelectric phase the curves $u_j(T)$ deviate from the quadratic law owing to the appearance of spontaneous polarization. However,  the curves $u_j(T)$ ($j=1,2,5$) cannot be  represented in the ferroelectric phase simply as a sum of  $u_j^{\text{para}}$ and of the item proportional to the spontaneous polarization, even roughly. Moreover, the differences $u_j-u_j^{\text{para}}$ ($j=1,2,5$) change their sign at some temperature (figure~\ref{Delta_u}), which qualitatively agree with experimental data  \cite{Deguchi3074}.

\begin{figure}[!t]
	\begin{center}
\includegraphics[scale=0.8]{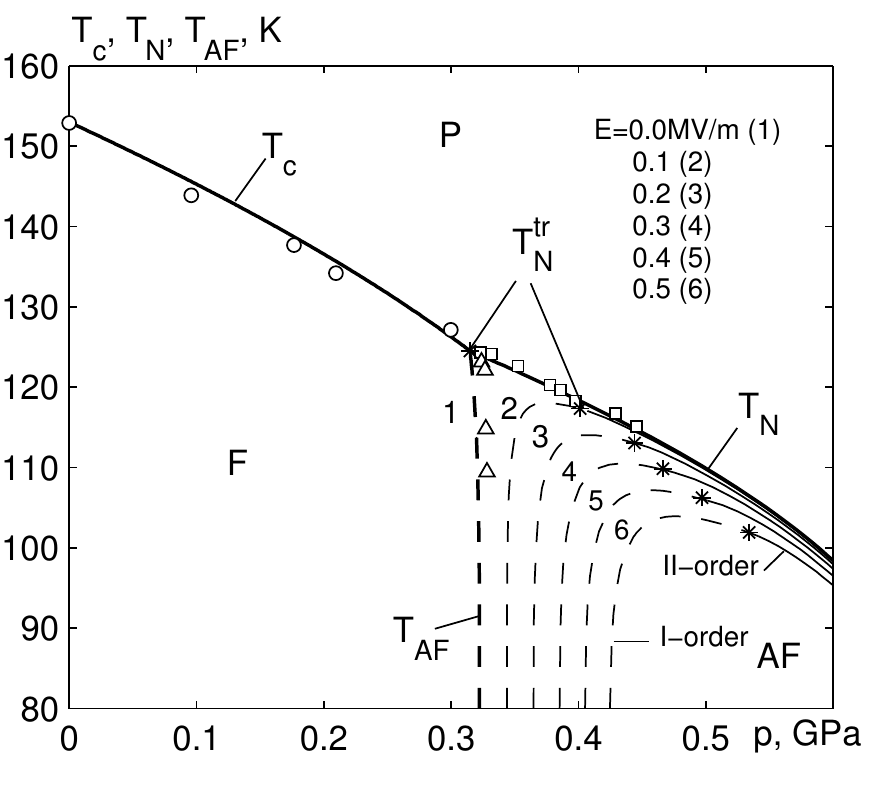} ~~ \includegraphics[scale=0.8]{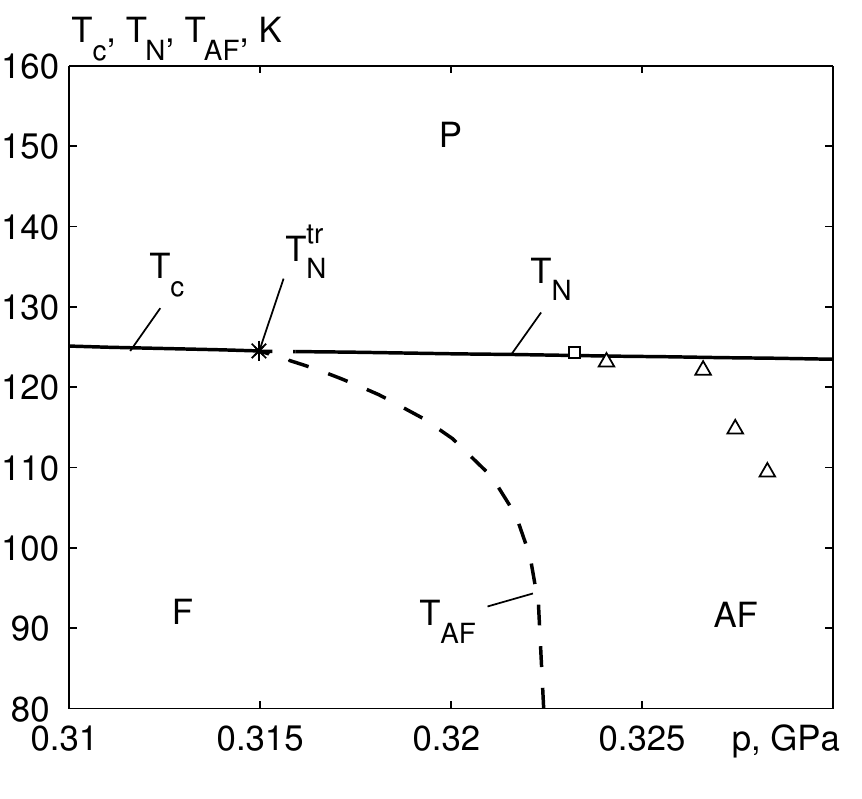} \\
		a ~~~~~~~~~~~~~~~~~~~~~~~~~~~~~~~~~~~~~~~~~~~~~~~~~~~~~~ b
	\end{center}
	\caption[]{(a) Pressure dependence of the  phase transition temperatures paraelectric-ferroelectric  ($T_c$),   paraelectric-antiferroelectric  ($T_\text{N}$) and ferroelectric-antiferroelectric ($T_{\text{AF}}$) of CDP crystal at different values of the electric field $E_2$(MV/m): 0.0 --1, 0.1 -- 2, 0.2 -- 3, 0.3 -- 4, 0.4 -- 5, 0.5 -- 6. Symbols are  experimental data, taken from \cite{Yasuda1311}. Tricritical points $ T_\text{N}^{\text{tr}} $ (denoted as *) separate the curves of first order (dashed lines) and second order  (solid lines) phase transitions. (b) The same phase diagram, but near critical pressure.} \label{TcTN_E2}
\end{figure}

\begin{figure}[!t]
\begin{multicols}{2}
\begin{center}
\includegraphics[scale=0.8]{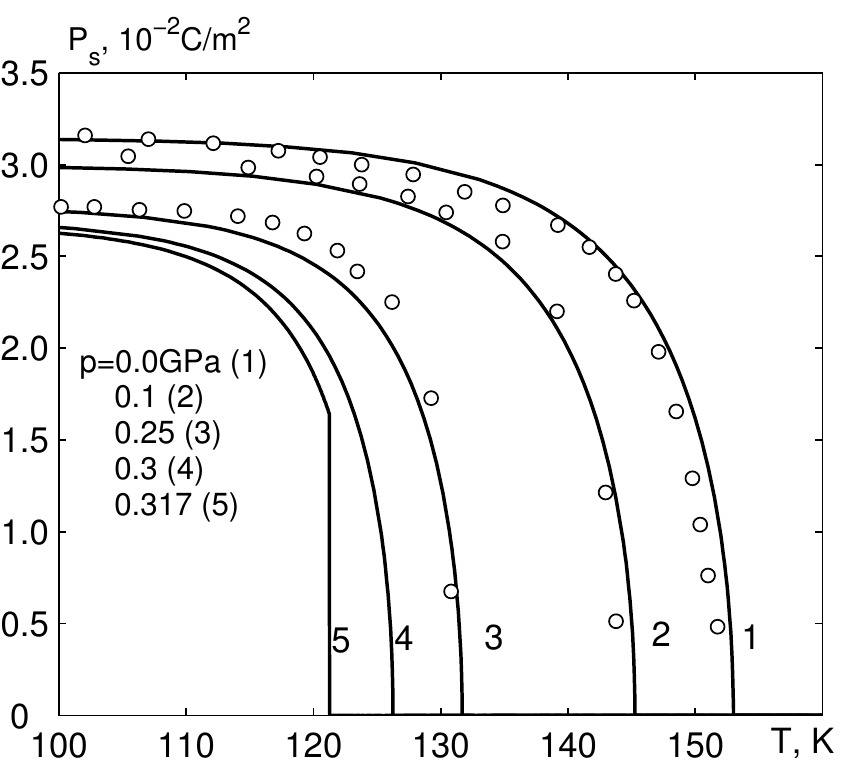} 
\end{center}
\caption[]{Temperature dependence of spontaneous polarization of  CDP at different values of hydrostatic pressure $p$, GPa: 0.0 -- 1, 0.1 -- 2, 0.25 -- 3, 0.3 -- 4, 0.317 -- 5. Symbols $\circ$ are experimental data \cite{Yasuda1311}.} \label{Psp_CDP_Yasuda}
\begin{center}
\includegraphics[scale=0.8]{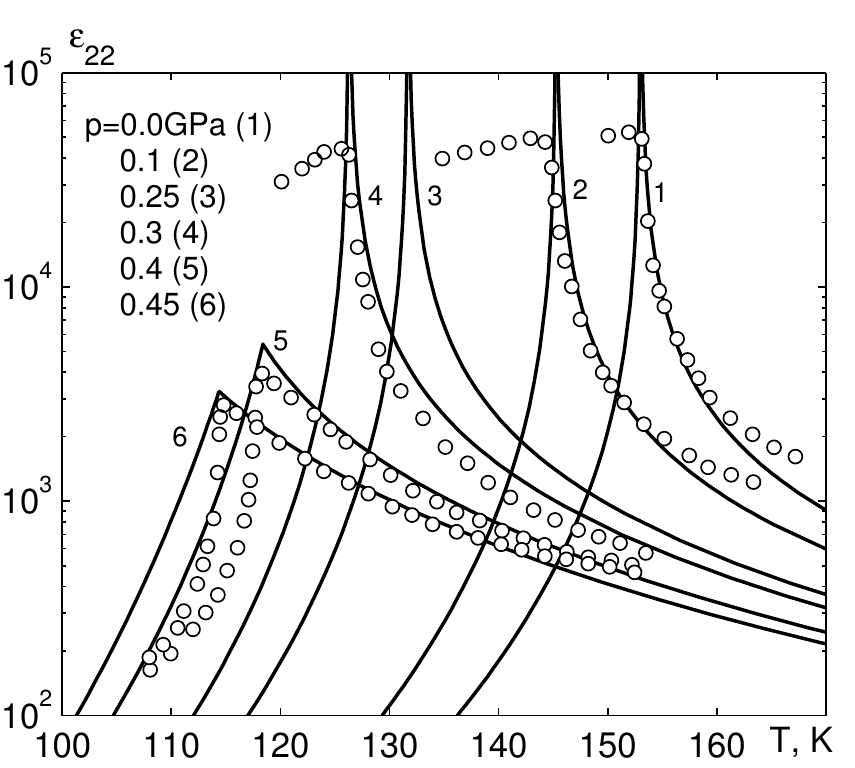}
\end{center}
\caption[]{Temperature dependence of  longitudinal dielectric permittivity  of  CDP at different values of hydrostatic pressure $p$,~GPa: 0.0 -- 1, 0.1 -- 2, 0.25 -- 3, 0.3 -- 4, 0.4 -- 5, 0.45 -- 6. Symbols $\circ$ are experimental data of \cite{Yasuda1311}.} \label{e22_CDP_Yasuda}	
\end{multicols}
\end{figure}

The lattice strains $u_j$  practically linearly depend on pressure according to Hooke's law.
According to (\ref{w}), (\ref{nu1}), (\ref{nu2}), this leads to a linear weakening under pressure of  interaction parameters $w$, $\nu_1$, $\nu_2$, respectively.  Here, in the range of pressure $0\,\,\text{GPa}<p<0.6\,\,\text{GPa}$, the parameter of short-range interactions  $w$ decreases  by 10\% (from 657~K to 602~K), whereas the parameters of long-range interactions  $\nu_1$, $\nu_2$ decrease up to negative values  (figure~\ref{nu_p}).
Consequently, the phase transition temperature lowers  (see figure~\ref{TcTN_E2}, curve $T_c$). Such dependence  $T_c(p)$ exists up to some critical pressure  $p_c$.
Here, the phase transition at the $T_c$ point remains to be the second order  transition, and the temperature dependences of different thermodynamic characteristics do not qualitatively change under pressure. In particular, spontaneous polarization monotonously and continuously decreases with an increasing  temperature and tends to zero at the $T_c$  point (figure~\ref{Psp_CDP_Yasuda}, curves 1--4); dielectric permittivity  $\varepsilon_{22}$ tends to infinity at the temperature $T_c$ (figure~\ref{e22_CDP_Yasuda}, curves 1--4). 
It is necessary to note that the theory concerns the monodomain crystal; it does not take into account reorientation of domain walls, which gives a large contribution to the experimentally measured permittivity in ferroelectric phase. Therefore, the permittivity  $ \varepsilon_{22} $ does not agree with experimental data in ferroelectric phase.
\begin{figure}[!t]
		\begin{minipage}[h]{0.45\linewidth}
			\includegraphics[width=1\textwidth]{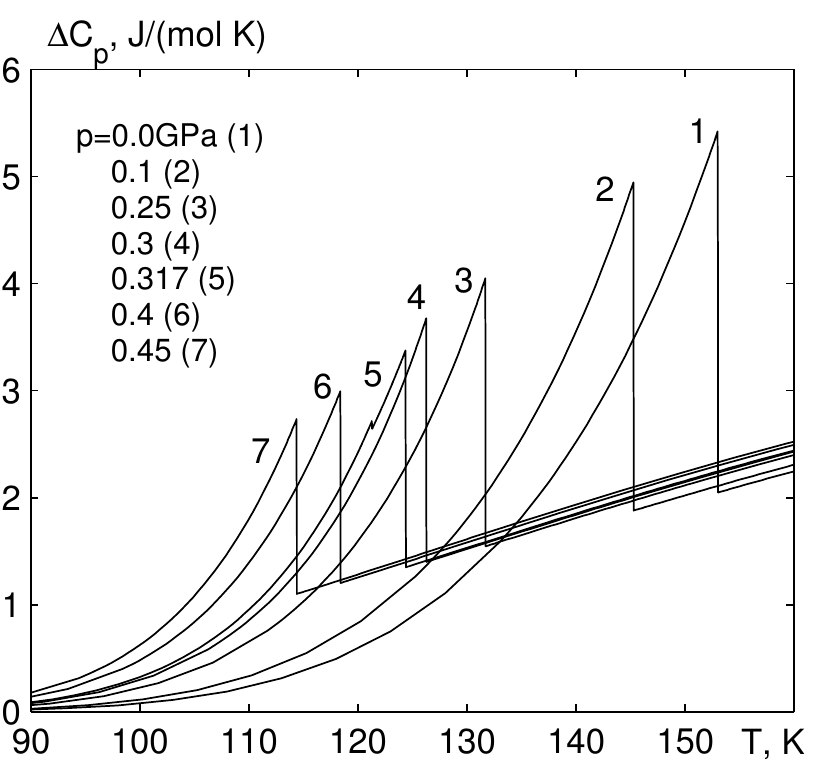}
			\caption{(Colour online) Temperature dependence of  the  proton contribution to molar heat capacity of  CDP at different values of hydrostatic pressure $p$,~GPa: 0.0 -- 1, 0.1 -- 2, 0.25 -- 3, 0.3 -- 4, 0.317 -- 5, 0.4 -- 6, 0.45 -- 7.} \label{DeltaC}
		\end{minipage}
		\hfill
				\begin{minipage}[h]{0.47\linewidth}
						\vspace{-2ex}
		\includegraphics[width=1\textwidth]{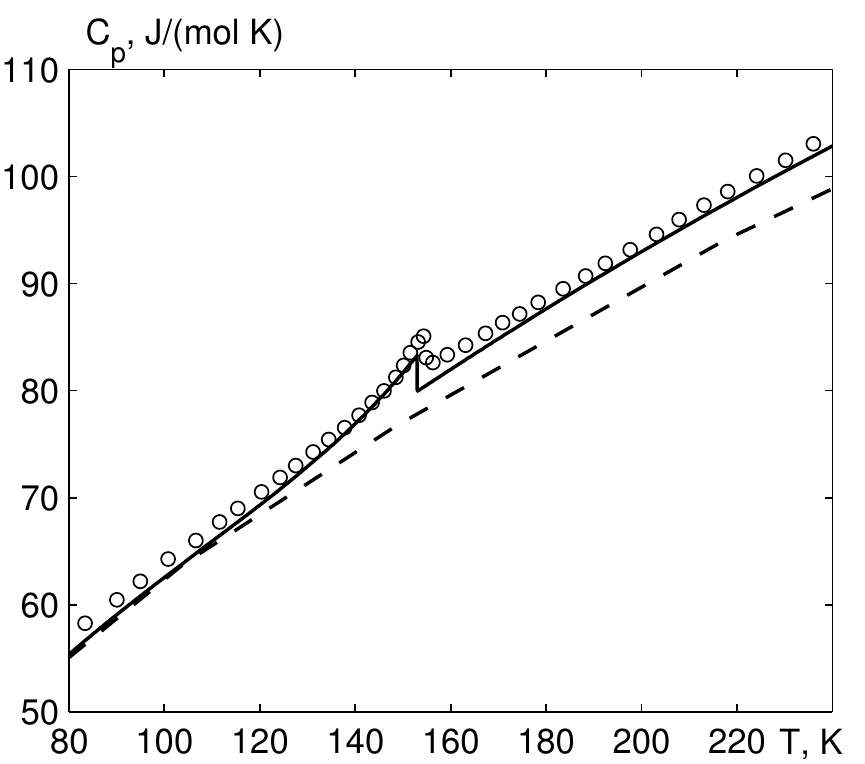}
			\caption{Temperature dependence of  the total  molar heat capacity of CDP. Symbols $\circ$ are experimental data taken from \cite{Imai3960}, dashed line is a result of ab-initio calculations  \cite{Shchur301}.} \label{C}
		\end{minipage}
\end{figure}
Temperature dependence of  the proton contribution to molar heat capacity also does not qualitatively  change under pressure  (figure~\ref{DeltaC}, curves 1--4). 
It has a jump  at the $T_c$ point, which slightly decreases with pressure. 
Total  molar heat capacity  (figure~\ref{C}, solid line) is the sum of the proton contribution and  the lattice contribution (dashed line) obtained in \cite{Shchur301} using ab-initio calculations.

\begin{figure}[!t]
	\begin{multicols}{2}
	\begin{center}
\includegraphics[scale=0.77]{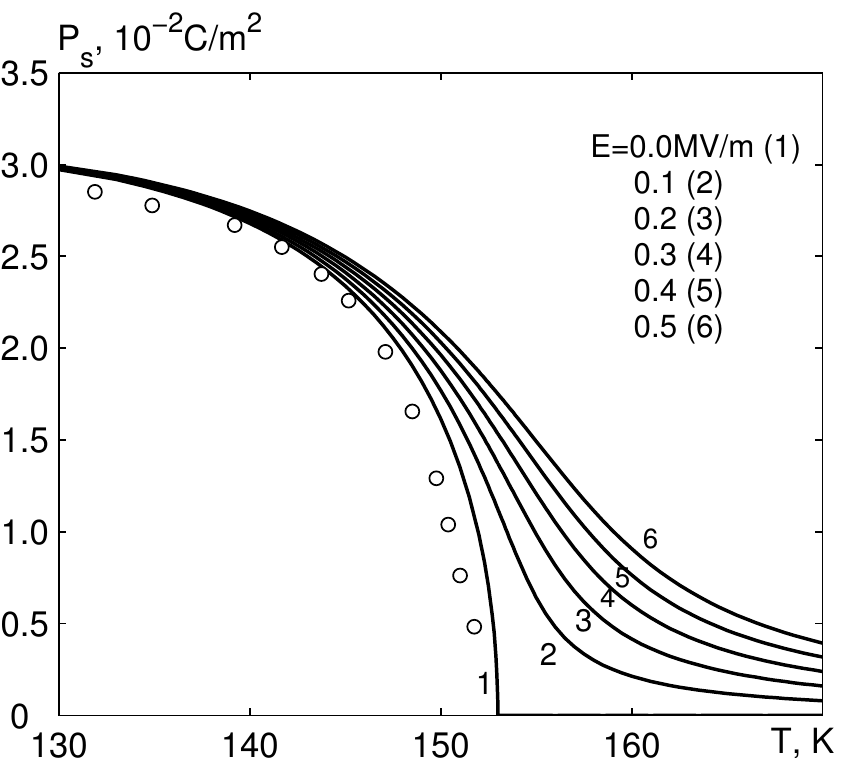} 
\end{center}
\caption{The temperature dependence of polarization of CDP crystal at $p$=0~GPa at different values of electric field  $E_2$~(MV/m): 0.0 --1, 0.1 -- 2, 0.2 -- 3, 0.3 -- 4, 0.4 -- 5, 0.5 -- 6.   Symbols $\circ$ are experimental data  \cite{Yasuda1311}.} \label{Ps_CDP_Yasuda_0kbar}
	\begin{center}
\includegraphics[scale=0.77]{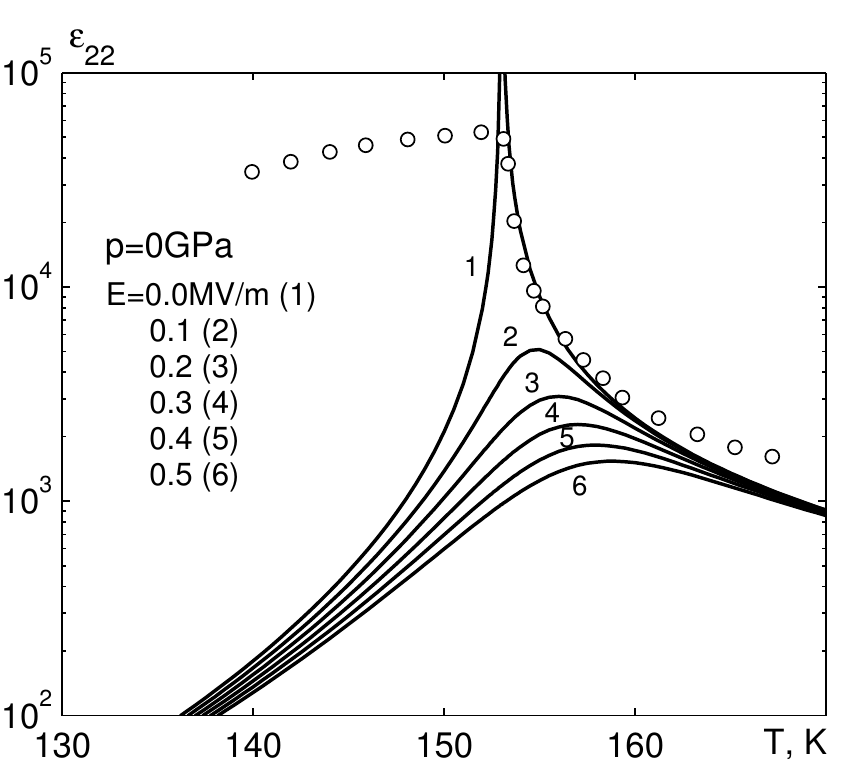}
\end{center}
\caption{Temperature dependence of dielectric permittivity of CDP crystal at $p$=0~GPa at different values of electric field  $E_2$~(MV/m): 0.0 --1, 0.1 -- 2, 0.2 -- 3, 0.3 -- 4, 0.4 -- 5, 0.5 -- 6. Symbols $\circ$ are experimental data \cite{Yasuda1311}.} \label{e22_CDP_Yasuda_0kbar}
	\end{multicols}
\end{figure}

\begin{figure}[!b]
	\begin{center}
\includegraphics[scale=0.77]{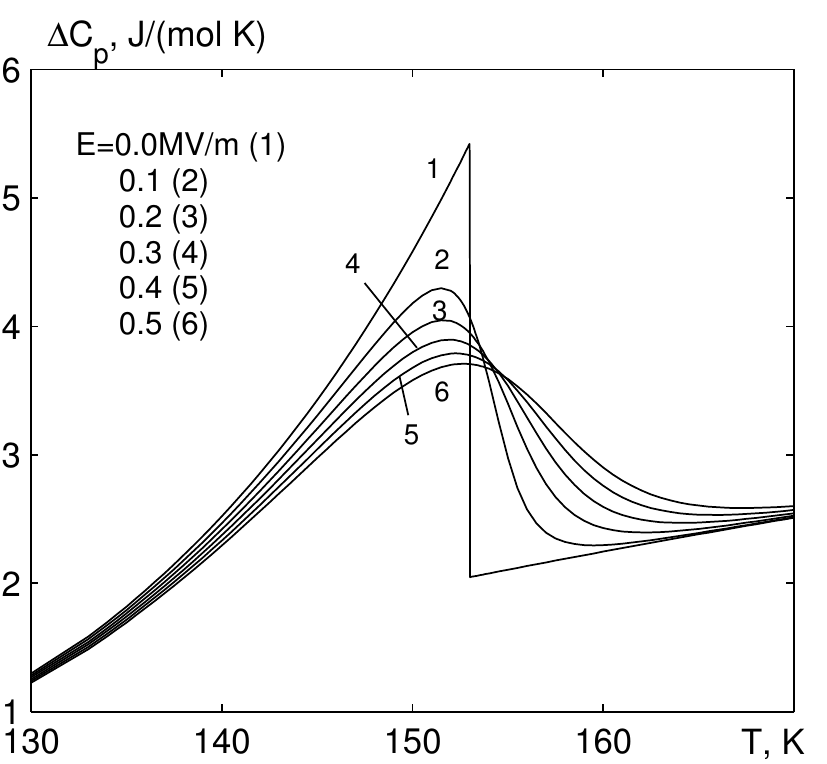}
	\end{center} 
	\caption{Temperature dependence of  the  proton contribution to molar heat capacity of  CDP at $p=0$~GPa at different values of electric field  $E_2$~(MV/m): 0.0 --1, 0.1 -- 2, 0.2 -- 3, 0.3 -- 4, 0.4 -- 5, 0.5 -- 6. } \label{DeltaC_0kbar}
\end{figure}

Longitudinal electric field  $E_2$ smears the phase transition from paraelectric to ferroelectric phase. As a result, curves $P_2(T)$, $\varepsilon_{22}(T)$ and $\Delta C_p(T)$ become smooth (see figures~\ref{Ps_CDP_Yasuda_0kbar},\ref{e22_CDP_Yasuda_0kbar},\ref{DeltaC_0kbar}, respectively).
In these figures, the field effect under zero pressure is demonstrated. Under  nonzero values of pressure  $p<p_c$, the  field effect is similar.


As it is shown above, the constants of long-range interactions  $\nu_{1}$ and $\nu_{2}$ linearly weaken with pressure. Starting from the critical pressure  $p_c=0.315$~GPa (experimental value is  $p_c=0.33\pm0.02$~GPa), the parameter of ``inter-sublattice'' interactions $\nu_{2}$ becomes negative (figure~\ref{nu_p}). Consequently, the sublattices ``A'' and ``B''  orient in opposite directions, and the crystal passes to antiferroelectric phase  (curve $T_\text{N}$ in figure~\ref{TcTN_E2})  instead of ferroelectric phase. Here,  spontaneous polarization is absent, and longitudinal permittivity  $\varepsilon_{22}$ is finite and has a sharp band in the $T_c$ point (figure~\ref{e22_CDP_Yasuda}, curves 5,6).

\begin{figure}[!b]
	\begin{center}
\includegraphics[scale=0.8]{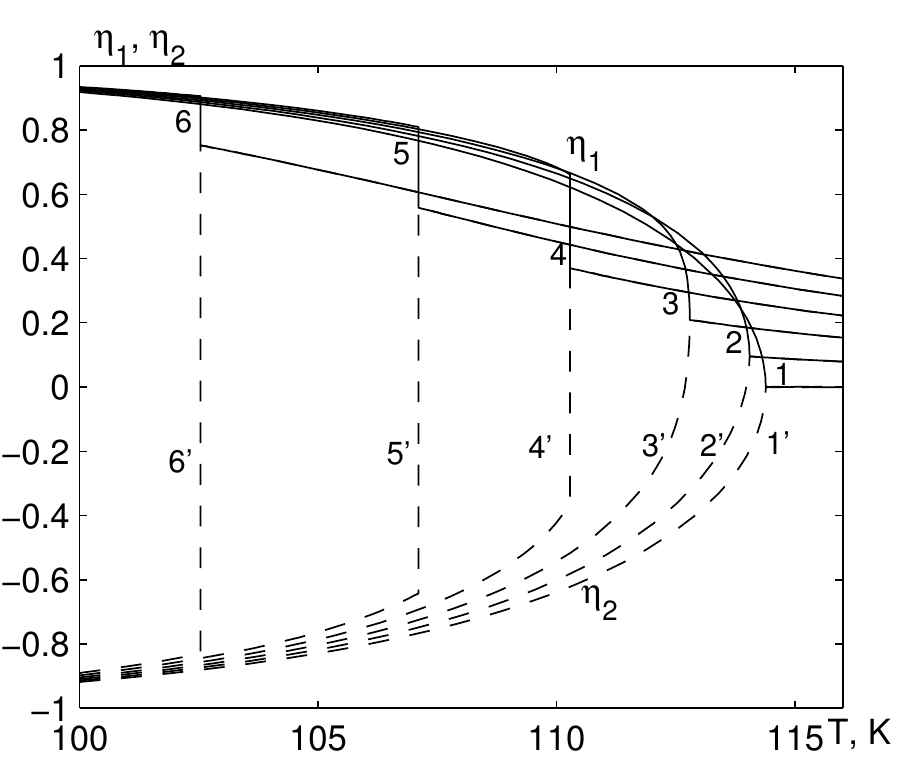}
\end{center}
\caption{Temperature dependences of the order parameters   $\eta_{1}$ (solid lines) and $\eta_{2}$ (dashed lines) at $p=0.45$~GPa at different values of electric field  $E_2$~(MV/m): 0.0 --1, 0.1 -- 2, 0.2 -- 3, 0.3 -- 4, 0.4 -- 5, 0.5 -- 6.} \label{eta_p4_5kbar}
\end{figure}

\begin{figure}[!b]
	\begin{multicols}{2}
	\begin{center}
\includegraphics[scale=0.75]{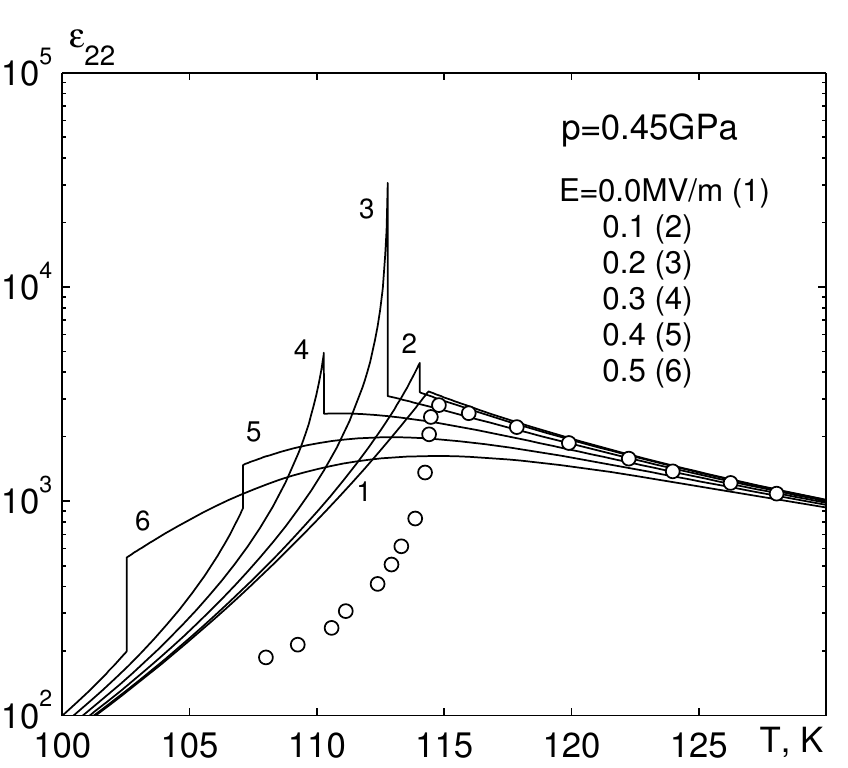}\\
\end{center}
\caption{Temperature dependence of dielectric permittivity of CDP crystal at $p=0.45$~GPa at different values of electric field  $E_2$~(MV/m): 0.0 --1, 0.1 -- 2, 0.2 -- 3, 0.3 -- 4, 0.4 -- 5, 0.5 -- 6. Symbols $\circ$ are experimental data  \cite{Yasuda1311}.} \label{e22_CDP_Yasuda_4_5kbar}
	\begin{center}
\includegraphics[scale=0.75]{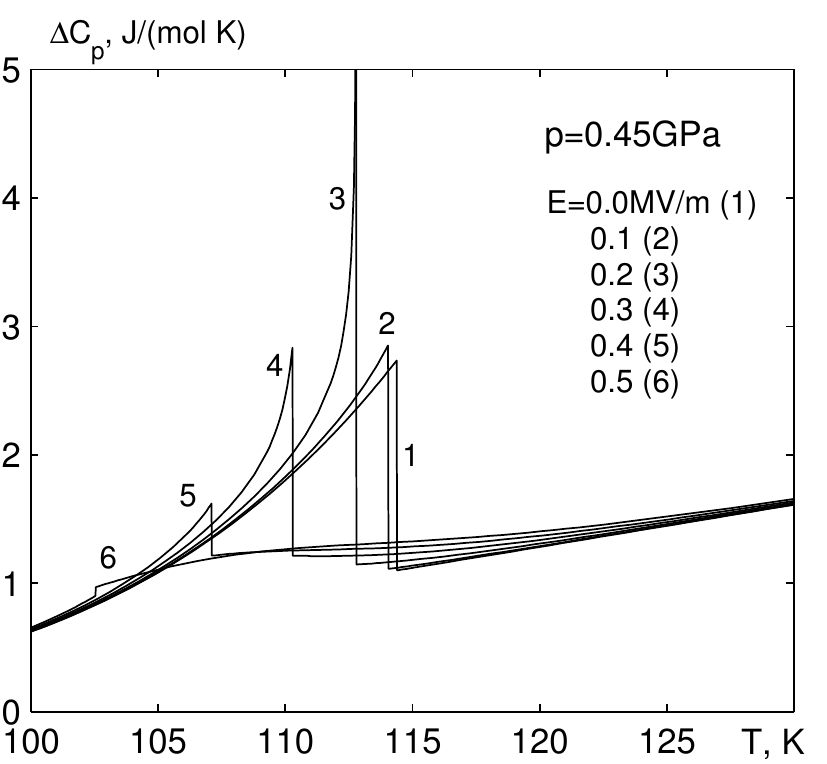}\\
\end{center}
\caption[]{Temperature dependence of the proton contribution to molar heat capacity of CDP crystal at $p=0.45$~GPa at different values of electric field  $E_2$~(MV/m): 0.0 --1, 0.1 -- 2, 0.2 -- 3, 0.3 -- 4, 0.4 -- 5, 0.5 -- 6. } \label{DeltaC_4kbar}
	\end{multicols}
\end{figure}

In the presence of electric field  $E_2$, the order parameter  $\eta_{1}$ slightly increases, in comparison with the case of  $E_2=0$ (figure~\ref{eta_p4_5kbar}, curves 1--6).  %
The parameter $\eta_{2}$, which is negative, on the contrary, decreases in value in comparison with the case of $E_2=0$ (figure~\ref{eta_p4_5kbar}, curves 1'--6'), moreover, a decrease in value of $\eta_{2}$ is stronger than an increase of $\eta_{1}$. That is, in antiferroelectric phase at the presence of the field $E_2$, the disordering of pseudospins in the sublattice ``A'' is stronger than the ordering in the sublattice~``B''. Consequently, in the antiferroelectric phase, dielectric permittivity  increases in comparison  with the case of $E_2=0$, and there appears a break at the $T_\text{N}$ point on the curves $\varepsilon_{22}(T)$ (figure~\ref{e22_CDP_Yasuda_4_5kbar}).
Molar heat capacity also increases in the antiferroelectric phase in comparison  with the case of $E_2=0$ (figure~\ref{DeltaC_4kbar}).
Inasmuch as the field induces polarization $P_2$ above $T_\text{N}$ point, then herein below the temperature region above $T_\text{N}$ in the presence of the field will be referred to as ``ferroelectric phase''. 
At  some critical value of the field $E_{\text{cr}}$ at constant values of pressure $p>p_c$ and temperature, the order parameter $\eta_{2}$ becomes positive, which means that there takes place an overturn of pseudospins in the sublattice ``B'', and the crystal passes from antiferroelectric to ferroelectric phase. 
The calculated values of $E_{\text{cr}}$ are several times larger in magnitude than the values experimentally measured in \cite{Yasuda2755}. In particular, at ${p}=0.38$~GPa at low temperatures $E_{\text{cr}}\approx 0.28~$MV/m, whereas the experimental value is $E_\text{cr}\approx 0.06$~MV/m. That is, the present model only qualitatively describes the field effect. 

Disagreement of the calculated $E_{\text{cr}}$ with experimental data can be explained in such a way. In expression (\ref{y2x}), the terms, that describe the interaction of a pseudospin with external field ${1}/{2}\beta(\mu_yE_2)$ and with the other sublattice $\beta \nu_{2}\eta_{1}$, have opposite signs (because at $p>p_c$ parameter $\nu_{2}<0$), which means that they compete with each other.  At the field $E>E_{\text{cr}}$, the interaction with external field prevails, and the order parameter $\eta_{2}$ changes its sign to the opposite. In order to obtain the value of  $\mu_yE_2$ large enough to turnover the pseudospin, at the weaker field, the dipole moment $\mu_y$ should be larger. On the other hand,  $\mu_y$ cannot be larger, because it fixes the saturation polarization, inasmuch as polarization depends mainly on the product $\mu_y(\eta_{1}+\eta_{2})$ [see (\ref{P2}), but the order parameters $\eta_{1},\eta_{2}\rightarrow 1$ when $T\rightarrow 0$].  

However, if we took into account the tunneling of protons on the hydrogen bonds, then the order parameters would be $\eta_{1},\eta_{2}<1$ at $T\rightarrow 0$. That is why the parameter $\mu_y$ in this case 
would be larger than without taking into account the tunneling. Strong isotopic  effect in the CDP may be the evidence of tunneling effects.  Besides, the distribution function of the proton momentum, obtained by ab-initio calculations in \cite{Lasave134112}, has  an additional peak at nonzero momentum at low temperatures, which points to a tunneling effect. It is also shown in \cite{Lasave134112}, that the shortening of the hydrogen bond with double-well potential by 1\% lowers the energy barrier between the equilibrium positions  several times. As a result, the tunneling effect on these bonds would greatly increase with pressure. Hence, the disagreement of $E_{\text{cr}}$ with the experimental data is connected, to a great extent, with neglecting the proton tunneling processes in the present model.

\begin{figure}[!t]
	\begin{center}
\includegraphics[scale=0.75]{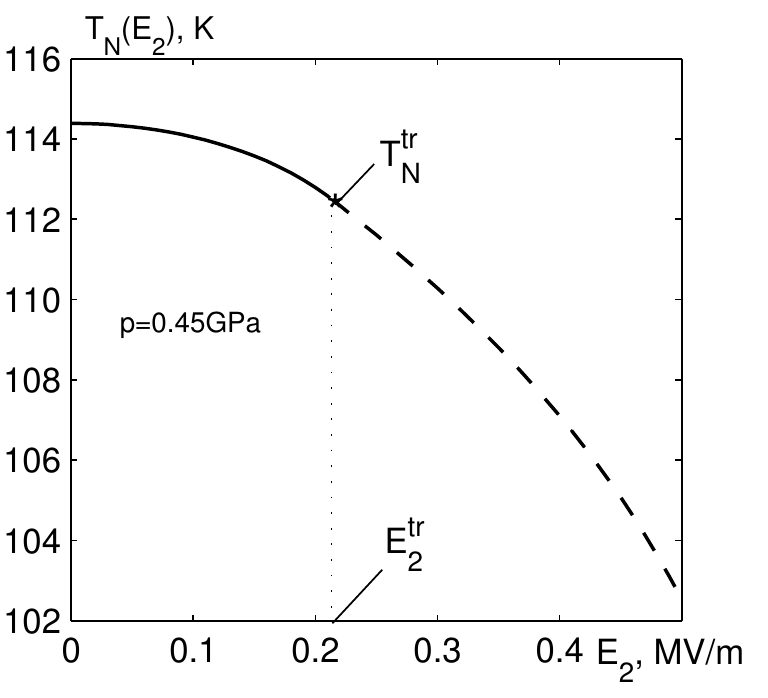} ~~~~~~~	\includegraphics[scale=0.75]{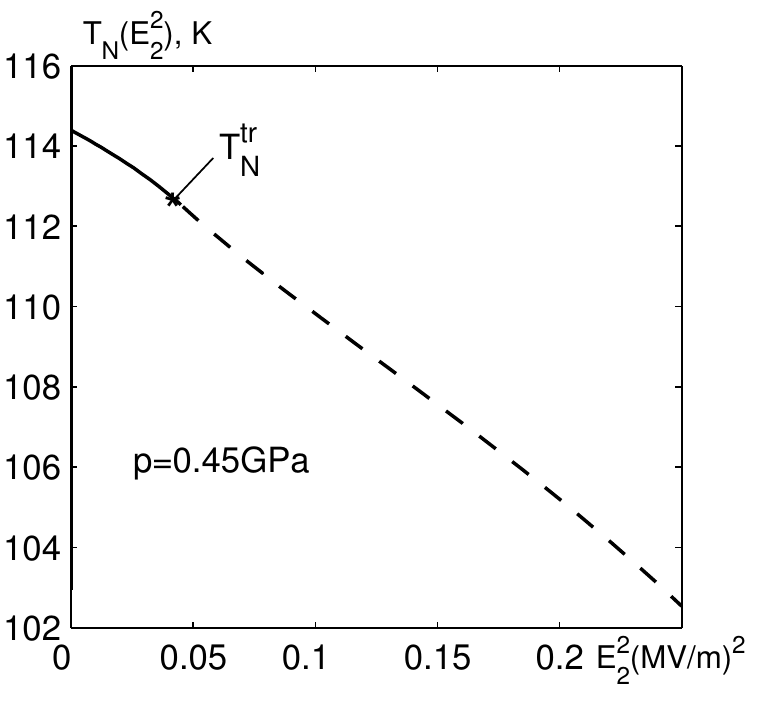}\\
		a ~~~~~~~~~~~~~~~~~~~~~~~~~~~~~~~~~~~~~~~~~~~~~~~~~~~~~~~~~~~~ b
	\end{center}
	\caption[]{Dependence of temperature $T_{\text{N}}$ on electric field $E_2$ (a) and on square of the field (b) at  $p=0.45$~GPa. Tricritical points  $ T_N^{\text{tr}} $ separate the curves of first order (dashed lines) and second order (solid lines) phase transitions.} \label{TN_E2}
\end{figure}

As a consequence of turnover of pseudospins at the presence of the external field at a constant pressure, the temperature  $T_\text{N}$ lowers approximately by the law  $T_\text{N}\sim -E_2^2$ (figure~\ref{TN_E2}). %
The lowering of $T_\text{N}$ also reveals itself  in the shift of the break on curves $\varepsilon_{22}(T)$ to lower temperatures (figure~\ref{e22_CDP_Yasuda_4_5kbar}), as well as in the suppression of the antiferroelectric region (AF) on the phase diagram (see figure~\ref{TcTN_E2}, curves 2--6). As one can see from figure \ref{TcTN_E2},  the closer is the value of hydrostatic pressure to the critical value $p_c$, the stronger is the effect of the field $E_2$ on the temperature $T_{\text{N}}$.

It is necessary to note that in weak fields, the phase transition in the $T_{\text{N}}$ point remains the second order phase transition (solid lines in figure~\ref{TN_E2} and in figure~\ref{TcTN_E2}, curves 2--6), but starting from some value of the field $E_2^{\text{tr}}$ (tricritical point) it becomes a first order phase transition (dashed lines in figure~\ref{TN_E2} and in figure~\ref{TcTN_E2}, curves 2--6). 

The above mentioned increase of permittivity in antiferroelectric phase in comparison with the case of  $E_2=0$ takes place at fields  $E_2<E_2^{\text{tr}}$ (figure~\ref{e22_CDP_Yasuda_4_5kbar}, curves 2, 3). At the fields  $E_2>E_2^{\text{tr}}$, the permittivity $\varepsilon_{22}$ decreases again (figure~\ref{e22_CDP_Yasuda_4_5kbar}, curves 4--6), because the order parameters  $\eta_{1}$,  $\eta_{2}$ in antiferroelectric phase near $T_{\text{N}}$ temperature become closer to saturation at a further strengthening of the field  (see figure~\ref{eta_p4_5kbar}, curves 4, 4', 5, 5', 6, 6').

\begin{figure}[!t]
		\begin{minipage}[h]{0.45\linewidth}
\includegraphics[width=1\textwidth]{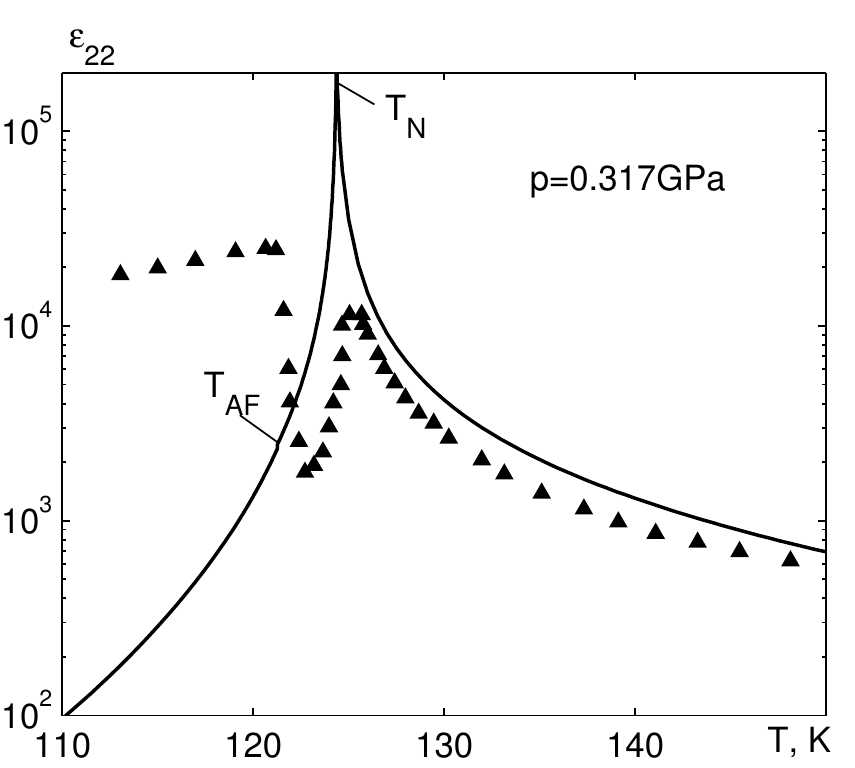}
			\caption{Temperature dependence of longitudinal dielectric permittivity of CDP at ${p}=0.317$~GPa. Symbols $\blacktriangle$ are experimental data~\cite{Yasuda1311}.} \label{e22_CDP_Yasuda_3_17kbar}
		\end{minipage}
		\hfill
				\begin{minipage}[h]{0.45\linewidth}
\includegraphics[width=1\textwidth]{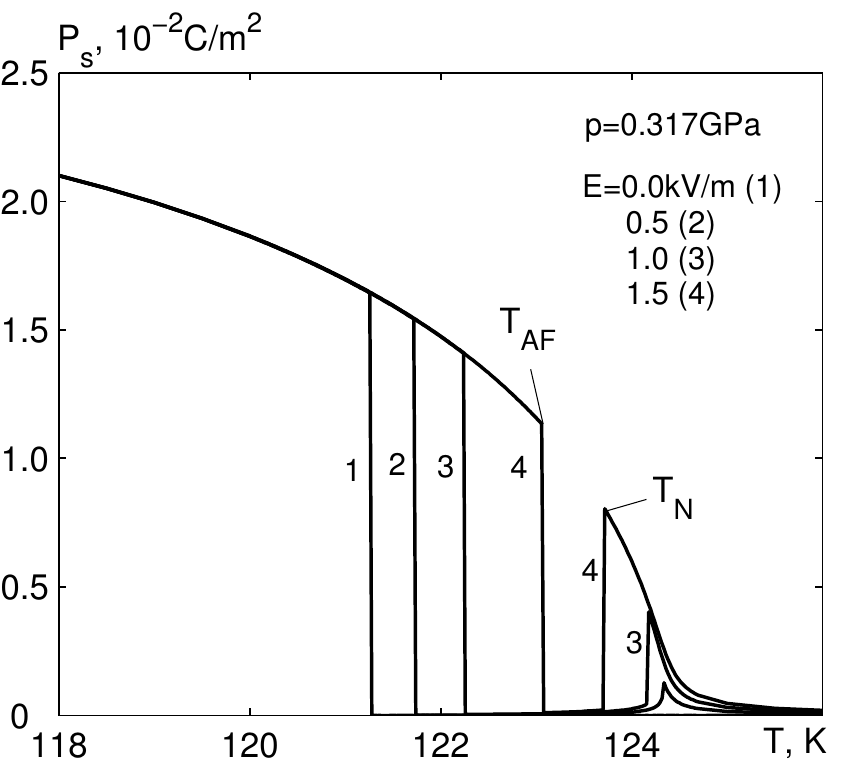}
			\caption[]{Temperature dependence of polarization of CDP at $p=0.317$~GPa at different values of electric field  $E_2$(kV/m): 0.0 --1, 1.0 -- 2, 2.0 -- 3, 3.0 -- 4. } \label{Ps_CDP_Yasuda_3_12kbar}
		\end{minipage}
\end{figure}

Let us investigate the behavior of  thermodynamic characteristics under pressures close to the critical: $p=0.315\div0.322$~GPa. 
Dependence of the lattice strains on temperature $u_j(T)$ reveals itself especially strongly under such pressures. Namely, the parameter of inter-sublattice interactions  $\nu_{2}$ changes its sign with the lowering of temperature. As a result, at a constant pressure, the crystal passes firstly from paraelectric to antiferroelectric phase at the temperature $T_{\text{N}}$, and with a further lowering of temperature  it passes from antiferroelectric to ferroelectric phase at some temperature $T_\text{AF}$, as one can see in figure~\ref{TcTN_E2}, b. Here, the phase transition at the  $T_\text{AF}$ point is the first order. In particular, at $p=0.317$~GPa, the spontaneous polarization exists under temperature  $T_\text{AF}=121.3$~K (figure~\ref{Psp_CDP_Yasuda}, curve 5), the temperature dependence of the proton contribution to the heat capacity $\Delta C_p$ has breaks at  temperatures  $T_\text{N}=124.4$~K and $T_\text{AF}=121.3$~K (see figure~\ref{DeltaC}, curve 5), and the temperature dependence of dielectric permittivity $\varepsilon_{22}$ has a sharp band at the  $T_\text{N}$ point and a small break at the  $T_\text{AF}$ point (see figure~\ref{e22_CDP_Yasuda_3_17kbar}).
As one can see from this figure, the agreement of theory with experimental data is only qualitative. Probably, the properties of CDP near the critical pressure  $p_c$ strongly depend on the quality of a sample. It is necessary to note that when the temperature  lowers the theoretical curve $\varepsilon_{22}$ has a jump down in the $T_\text{AF}$ point, whereas the experimental values of $\varepsilon_{22}$, on the contrary, abruptly increase due to the contribution to permittivity from reorientation of domain walls.

Near the critical pressure, the properties of a crystal are very sensitive to the electric field. From the temperature dependence of polarization at $p=0.317$~GPa (see figure~\ref{Ps_CDP_Yasuda_3_12kbar}) one can see that weak fields (up to 2~kV/m) greatly increase the temperature  $T_\text{AF}$ and  greatly lower the temperature $T_\text{N}$. 
%
Here, in the external field the temperature position of the break on the curves of dielectric permittivity and heat capacity at $T_\text{AF}$ point shifts to the  higher temperatures, whereas the effect of the field on the thermodynamic characteristics  near   $T_\text{N}$ point at $p=0.317$~GPa is qualitatively similar to the case of $p=0.45$~GPa. Consequently, at a constant pressure, the electric field $E_{2}$ decreases the temperature range, at which the antiferroelectric phase exists, up to its complete its disappearance, and at a constant temperature  this field increases the critical pressure   (see figure~\ref{TcTN_E2}, curves 2--6).

\section{Conclusions}

The lowering of the phase transition temperature $T_c$ with pressure is connected with the weakening of long-range and short-range (to a lesser extent) interactions. Under the pressures higher than some critical pressure $p_c$, the inter-sublattice interactions become negative. Consequently, there appear  paraelectric-antiferroelectric and ferroelectric-antiferroelectric phase transitions.

The dependence of the lattice strains (as well as long-range interactions) on temperature reveals itself mainly near the critical pressure. Namely,  the inter-sublattice interactions change their sign with the lowering of temperature. As a result, at a constant pressure, close to the critical pressure, the crystal passes first from paraelectric to  antiferroelectric phase, and with a further lowering of temperature it passes from antiferroelectric to ferroelectric phase.

Longitudinal electric field $E_{2}$ increases the critical pressure.
Under pressures $p<p_c$, the external field smears the phase transition.
Under pressures $p>p_c$, the external field lowers the temperature $T_\text{N}$ and increases the permittivity $\varepsilon_{22}$ in the antiferroelectric phase. This can be explained by a larger disordering of pseudospins in the sublattice  ``B'' than the ordering  in the sublattice ``A'' in the presence of the electric field. A strong enough field can change the order of phase transition at the $T_\text{N}$ point from second order to first order. The strongest field effect on the calculated characteristics takes place near the critical pressure.

%
%

\ukrainianpart

\title{Вплив гідростатичного тиску та поздовжнього електричного поля  на фазові переходи та термодинамічні характеристики квазіодновимірного сегнетоелектрика CsH$_2$PO$_4$}
\author{ А.С. Вдович \refaddr{label1}, І.Р. Зачек \refaddr{label2}, Р.Р. Левицький \refaddr{label1}}
\addresses{
	\addr{label1} Інститут фізики конденсованих систем НАН України, вул. Свєнціцького, 1,
	79011 Львів, Україна
	\addr{label2} Національний університет ``Львівська політехніка'', вул.~С.~Бандери,  12, 79013 Львів, Україна}

%
%
%

\makeukrtitle

\begin{abstract}
\tolerance=3000%
Запропоновано двопідграткову модель протонного впорядкування квазіодновимірного сегнетоелектрика з водневими зв'язками CsH$_2$PO$_4$, яка враховує  лінійні за деформаціями гратки $u_1$, $u_2$, $u_3$ і $u_5$ внески в енергію протонної підсистеми. Модель враховує також залежність ефективних дипольних моментів псевдоспінів від параметрів впорядкування, що дозволяє узгодити ефективні дипольні моменти в сегнето- і парафазі. У рамках цієї моделі  в наближенні двочастинкового кластера за короткосяжними і середнього поля за далекосяжними взаємодіями,  досліджено поведінку спонтанної поляризації, поздовжньої діелектричної проникності і молярної теплоємності під дією гідростатичного тиску і поздовжнього електричного поля. Пояснено перехід в антисегнетофазу при високих тисках. Вивчено характер розмиття фазового переходу парафаза-сегнетофаза, а також пригнічення антисегнетофази в електричному полі.
\keywords сегнетоелектрики, діелектрична проникність, фазові переходи, вплив гідростатичного тиску,  вплив електричного поля

\end{abstract}


\begin{thebibliography}{99}

	
\bibitem{Matsunaga2011} Matsunaga H., Itoh K., Nakamura E., J. Phys. Soc. Jpn., 1980, \textbf{48}, No.~6, 2011--2014, \doi{10.1143/JPSJ.48.2011}.  

\bibitem{Itoh2626} Itoh K., Hagiwara T., Nakamura E., J. Phys. Soc. Jpn., 1983, \textbf{52}, No.~8, 2626--2629, \doi{10.1143/JPSJ.52.2626}.       

\bibitem{Iwata304} Iwata Y., Koyano N., Shibuya I., J. Phys. Soc. Jpn., 1980, \textbf{49}, No.~1, 304--307, \doi{10.1143/JPSJ.49.304}.

\bibitem{Iwata4044} Iwata Y., Deguchi K., Mitani S., Shibuya I., Onodera Y., Nakamura E., J. Phys. Soc. Jpn., 1994, \textbf{63}, No.~11, 4044--4050, \doi{10.1143/JPSJ.63.4044}.

\bibitem{Yasuda1311} Yasuda N., Okamoto M., Shimizu H., Fujimoto S., Yoshino K., Inuishi Y., Phys. Rev. Lett., 1978, \textbf{41}, No.~19, 1311--1314, \doi{10.1103/PhysRevLett.41.1311}.

\bibitem{Yasuda2755} Yasuda N., Fujimoto S., Okamoto M., Shimizu H., Yoshino K., Inuishi Y.,  Phys. Rev. B., 1979, \textbf{20}, No.~7, 2755--2764, \doi{10.1103/PhysRevB.20.2755}.

\bibitem{Schuele935} Schuele P.J.,  Thoma R.A., Jpn. J. Appl. Phys., 1985, \textbf{24}, 935--937,  \doi{10.7567/JJAPS.24S2.935}.

\bibitem{Schuele2549} Schuele P.J., Schmidt V.H.,   Phys. Rev. B., 1989, \textbf{39}, No.~4, 2549--2556,  \doi{10.1103/PhysRevB.39.2549}.

\bibitem{Kobayashi83} Kobayashi Yu., Deguchi K., Azuma Sh., Suzuki E., Ming Li Ch., Endo Sh., Kikegawad T., Ferroelectrics, 2003, \textbf{285}, No.~7, 83--89,  \doi{10.1080/00150190390205924}.

\bibitem{Brandt} Brandt N.B., Zhukov S.G., Kulbachinskii V.A., Smirnov P.S., Strukov B.A., Fiz. Tverd. Tela, 1986, \textbf{28}, 3159, (in Russian).

\bibitem{Magome2010} Magome E., Tomiaka S., Tao Y., Komukae M., J. Phys. Soc Jpn., 2010, \textbf{79}, No.~2,  025002, \\ \doi{10.1143/JPSJ.79.025002}.

\bibitem{Blinc6031} Blinc R., SaBaretto F.C., J. Chem. Phys., 1980, \textbf{72}, No.~11, 6031--6034, \doi{10.1063/1.439058}.

\bibitem{914R} Stasyuk I.V., Levytsky R.R., Zachek I.R.,  Shchur Ya.Y., Kutny J.V., Miz E.V.,  Preprint of the Institute for Condensed Matter Physics, ICMP–91–4R, Lviv,
1991, (in Russian).

\bibitem{Deguchi3074} Deguchi K., Okaue E., Ushio S., Nakamura E., Abe K., J. Phys. Soc. Jpn, 1984, \textbf{53}, No.~9, 3074--3080,\\ \doi{10.1143/JPSJ.53.3074}.

\bibitem{FXTT40} Levitskii R.R., Zachek I.R., Vdovych A.S.  Phys. Chem. Solid State, 2012, \textbf{13}, No.~1, 40--47.

\bibitem{Levitskii4702} Levitskii R.R., Zachek I.R., Vdovych A.S., J. Phys. Stud., 2012, \textbf{16}, No.~4, 4702 (in Ukrainian).


\bibitem{Imai3960} Imai K., J. Phys. Soc. Jpn., 1983, \textbf{52}, No.~11, 3960--3965, \doi{10.1143/JPSJ.52.3960}.

\bibitem{Prawer63} Prawer S., Smith T.F.,  Finlayson T.R.,  Aust. J. Phys., 1985, \textbf{38}, No.~1, 63--83,  \doi{10.1071/PH850063}.

\bibitem{Shchur301} Shchur Ya., Bryk T., Klevets I., Kityk A.V.,  Comput. Mater. Sci., 2016, \textbf{111}, 301–309,\\ \doi{10.1016/j.commatsci.2015.09.014}.

\bibitem{VanTroeye024112} Van Troeye B., van Setten M.J., Giantomassi M., Torrent M., Rignanese G.-M.,  Gonze X.,  \\ Phys. Rev. B., 2017, \textbf{95}, No.~2, 024112,  \doi{10.1103/PhysRevB.95.024112}.

\bibitem{Lasave134112} Lasave J., Abufager P., Koval S., Phys. Rev. B, 2016, \textbf{93}, No.~13, 134112, \doi{10.1103/PhysRevB.93.134112}.

	
\end{thebibliography}
\end{document}